\documentclass[11pt]{article}

\usepackage{amsmath}
\usepackage{graphicx}
\usepackage{indentfirst}
\usepackage{amssymb}
\usepackage{cite}
\usepackage{color}
\usepackage{subfigure}
\usepackage{varwidth}
\usepackage{subfigure}
\usepackage[colorlinks=true, linkcolor=red, citecolor=blue, urlcolor=magenta]{hyperref}
\usepackage{xcolor}

\begin{document}
\title{Observational features of the Bardeen-boson star with thin disk accretion}

\date{}
\maketitle

\begin{center}
\author{Chen-Yu Yang,}$^{a}$\footnote{E-mail: chenyu\_yang2024@163.com}
\author{Huan Ye,}$^{b}$\footnote{E-mail: yehuan10053@163.com}
\author{Xiao-Xiong Zeng}$^{c}$\footnote{E-mail: xxzengphysics@163.com (Corresponding author)}
\\

\vskip 0.25in
$^{a}$\it{Department of Mechanics, Chongqing Jiaotong University, Chongqing 400000, People's Republic of China}\\
$^{b}$\it{School of Material Science and Engineering, Chongqing Jiaotong University, Chongqing 400074, People's Republic of China}\\
$^{c}$\it{College of Physics and Electronic Engineering, Chongqing Normal University, Chongqing 401331, People's Republic of China}\\

\end{center}
\vskip 0.6in
{\abstract
{In this work, we construct spherically symmetric solutions of Bardeen--boson stars within the framework of the Einstein--Klein--Gordon theory coupled to nonlinear electrodynamics (NLED) by employing numerical methods. Considering a thin accretion disk in the equatorial plane as the light source, we systematically investigate the optical appearance of boson stars using the ray-tracing method and the stereographic projection technique. Particular attention is paid to the influence of the initial scalar field $\phi_0$, the magnetic charge $\mathcal{G}$, and the observation angle $\theta_o$, on the image structure. As compact horizonless objects, boson stars produce optical images dominated by direct emission, while their morphology undergoes significant distortions as $\theta_o$ increases. Higher values of $\phi_0$ and $\theta_o$ can give rise to lensing images. For all the  parameters, the image center exhibits a brightness depression similar to the inner shadow of black holes, which poses challenges for distinguishing between boson stars and black holes. To address this, we propose two possible approaches: (i) combining the analysis of lensing bands with the effective potential to determine the existence of photon rings; and (ii) examining the polarization effects under synchrotron emission mechanisms. These results provide theoretical support for future high-resolution imaging efforts aimed at discriminating boson stars from black holes.
}}

\thispagestyle{empty}
\newpage
\setcounter{page}{1}


\section{Introduction}
One of the most important predictions of general relativity (GR) is the existence of black holes. Since the discovery of the first class of black hole solutions, namely the Schwarzschild black hole, the study of black hole solutions and their properties has attracted significant attention. However, for a long time, research on black holes remained primarily theoretical, a situation that has been remarkably improved in recent years. Gravitational waves generated by binary black hole mergers have been successfully detected~\cite{abbott2016observation}, and the Event Horizon Telescope (EHT) Collaboration has released images of the central black holes in M87 and Sagittarius A$^\ast$ at 230\,GHz~\cite{event2019first,akiyama2022first}, as well as their polarization images~\cite{akiyama2021first,akiyama2024first}. These groundbreaking observations confirmed the validity of GR and unveiled the physical mechanisms of accretion disk dynamics and jet formation, marking the advent of the observational era of compact object studies. Ultra-compact objects, such as black holes and boson stars, play a crucial role in probing extreme physical processes and testing theories of gravity. The shadows and optical features of these compact objects can be used to measure their masses with precision, while also revealing essential information about their spacetime structure, magnetic fields, and the distribution of accreting matter. Therefore, imaging studies of ultra-compact objects are instrumental for both distinguishing between different classes of astrophysical objects in practice and providing new insights into extreme gravitational environments and fundamental physics.

As typical ultra-compact objects, the shadow images of black holes have been extensively studied~\cite{wei2013observing,zeng2025kerr,hou2022image,wang2025geodesics,zhang2025autocorrelation}. In 1973, Shakura and Sunyaev proposed the earliest accretion disk model, the $\alpha$-disk, which describes a geometrically thin but optically thick standard accretion disk~\cite{shakura1973black}. Subsequently, in 1974, the Shapiro–Lightman–Eardley accretion disk model, characterized by both geometrical and optical thinness, was introduced; this model also incorporates the relativistic effects of matter motion near black holes~\cite{lightman1974black}. In 1979, Luminet employed a semi-analytical approach to draw the shadow of a Schwarzschild black hole, demonstrating that the shadow appears as a perfect circle to observers at any viewing angle~\cite{luminet1979image}. Since then, black hole shadows in Kerr spacetimes with Keplerian accretion disks have been investigated in detail~\cite{beckwith2005extreme}. In recent years, a wide variety of accretion models have attracted considerable attention, including spherical accretion models~\cite{narayan2019shadow,zeng2020shadows,heydari2023shadows}, optically and geometrically thin accretion disks~\cite{guo2024image,yang2024shadow,he2024observational,zeng2022shadows,li2021shadows,vincent2011gyoto}, and geometrically thick accretion disks~\cite{zhang2024images,zhang2024imaging,gjorgjieski2024comparison}. Moreover, based on the EHT observational data, numerous studies have attempted to compare theoretical predictions with observed black hole shadows~\cite{zeng2025holographic,cui2024optical,guo2024influence,hou2024unique,huang2024images}, constraining the parameters of different black hole models to ensure consistency between the predicted shadow sizes and those observed for M87* and Sgr~A*. Therefore, both observationally and theoretically, the shadows and optical imaging of black holes and other compact objects have increasingly become a focal point of research.

However, as the most distinctive feature of black holes, the event horizon still poses significant challenges for direct observational verification. Moreover, the observational uncertainties in current black hole imaging have spurred rapid development in the study of black hole “mimickers”~\cite{sathyaprakash2019extreme}. At present, the precision of EHT observations is not sufficient to completely rule out the possibility that the central object is a horizonless compact star rather than a black hole. Since horizonless ultra-compact objects share many properties with black holes, investigating their shadows and optical images becomes particularly important. Among the various black hole mimickers, boson stars occupy a prominent position due to their unique structure and physical properties. A boson star is a type of ultra-compact object composed of bosons bound by self-gravity. Since the pioneering works of Kaup~\cite{kaup1968klein}, Ruffini and Bonazzola~\cite{ruffini1969systems}, the stability and dynamical properties of boson stars have been extensively studied~\cite{liebling2023dynamical,di2020dynamical}. A variety of boson star solutions have been constructed, including charged boson stars~\cite{jetzer1993charged}, Newtonian boson stars~\cite{silveira1995boson}, and rotating multistate boson stars~\cite{li2020rotating}. In addition, gravitational wave signals from binary boson star systems under solitonic potentials have been systematically investigated~\cite{bezares2022gravitational}, and boson stars have even been considered as potential candidates for dark matter~\cite{pitz2025generating}. Some studies suggest that Sgr~A* at the Galactic center could in fact be a boson star~\cite{vincent2016imaging}. Following the release of black hole images by the EHT, the optical properties of boson stars have received increasing attention. Owing to the absence of an event horizon, light rays can traverse the interior of boson stars and reach the observer, leading to imaging features that differ significantly from those of black holes. Various studies have explored boson star optical images under different potentials and modified gravity frameworks~\cite{maso2021boson,rosa2023imaging,rosa2024accretion,he2025observation}, including mini boson stars~\cite{zeng2025opticalimagesminiboson} and massive boson stars~\cite{li2025observational}. It has been shown that under truncated accretion disks~\cite{herdeiro2021imitation} and for small observer inclination angles, boson stars can mimic the shadows of Schwarzschild black holes~\cite{rosa2022shadows}. On the other hand, the polarized images of different black hole models and horizonless compact objects have also been systematically investigated~\cite{angelov2025polarized,qin2022polarized,shi2024polarized,liu2022polarization,hu2022polarized,zhang2024polarized,wang2025semi,huang2024coport,hou2025near,zeng2025polarization}, demonstrating that polarization imaging provides valuable insights into the geometry of compact objects and may even serve to distinguish different classes of compact objects~\cite{deliyski2022polarized,deliyski2023polarized}. Against this background, the present work focuses on the optical images of Bardeen-boson stars.

The Bardeen model holds a milestone position in GR, as it provides the first regular (singularity-free) black hole solution~\cite{bardeen1968non}. In 1968, Bardeen introduced a de Sitter–like vacuum region at the black hole center to replace the classical singularity, thereby eliminating the divergent curvature structure geometrically. Although originally proposed as a heuristic construction, Ayón-Beato and García later demonstrated that this model can be consistently derived from nonlinear electrodynamics (NLED) and remains compatible with the Einstein equations, thus establishing a field-theoretic foundation for the existence of “regular black holes”~\cite{ayon1998regular}. However, NLED-based regular black holes do not always reduce to standard Maxwell theory in the weak-field limit. Studies have shown that if one requires the weak-field limit to recover the Maxwell form while preserving global regularity, such conditions are typically satisfied only in the purely magnetic case, whereas the electrically charged configurations often conflict with either regularity or the Maxwell limit~\cite{bronnikov2023regular}. Recent investigations have further revealed that, within the framework of GR, introducing a scalar field nonminimally coupled to the electromagnetic field through a coupling function $W(\phi)$ can self-consistently maintain a regular spacetime structure~\cite{cordeiro2025black}. The coupling term $W(\phi)L(F)$ effectively modulates the electromagnetic contribution to gravity, achieving a dynamical balance between geometry and matter sources. Analytical reconstruction of $L(r)$, $L_F(r)$, and $W(r)$ shows that, for specific parameter choices (e.g., $n=0$), one can still obtain a regular solution even without invoking traditional NLED. This implies that linear electrodynamics can also lead to a nonsingular geometry. Such a mechanism provides a new physical interpretation for Bardeen-type regular geometries: the scalar field modifies electromagnetic interactions through $W(\phi)$, ensuring spacetime regularity while retaining a linear electrodynamic form. Once the theoretical framework of Bardeen regular spacetimes is established, studying their observable features becomes an important avenue to understand their physical nature. In particular, the black hole shadow, as a direct probe of the strong-gravity geometry, offers an ideal platform to test the distinctions between regular and singular black holes.

Significant progress has been made in the study of the Bardeen black hole shadow. He \emph{et al.} systematically analyzed the shadow features of the Bardeen black hole under different accretion environments and found that the shadow radius remains nearly unchanged across various accretion modes, although the overall brightness is considerably reduced in the infalling accretion model. In the optically thin disk model, the observed emission is mainly dominated by direct radiation, while the contributions from the lensing and photon rings are negligible. Moreover, variations in the Bardeen model parameters, such as the magnetic charge, can significantly alter the shadow shape and brightness distribution~\cite{he2022shadow}. Furthermore, He \emph{et al.} investigated the optical appearance of rotating Bardeen black holes in a perfect fluid dark matter background. Their results show that as the observation angle increases, the shadow brightness becomes increasingly concentrated in the lower region of the image. Increasing the dark matter parameter enlarges this region, while a higher inclination enhances the redshift effect. In contrast, larger values of the magnetic charge, spin parameter, or dark matter content tend to weaken the redshift~\cite{he2025observational}. Stuchlík and Schee demonstrated that NLED effects can produce observable deviations of up to about 20\% in the shadow features of regular black holes and strongly influence the photon deflection angle in the Bardeen spacetime~\cite{stuchlik2019shadow}. Guo \emph{et al.} further pointed out that, although the Bardeen and Schwarzschild black holes appear morphologically similar, their brightness distributions differ significantly. This difference indicates that the regular spacetime structure of the Bardeen black hole can enhance the radiative efficiency of the accretion disk, an effect driven by the stronger curvature induced by the magnetic charge~\cite{guo2023unveiling}.

Building on these results, researchers have further explored self-consistent configurations obtained by introducing a complex scalar field into the Bardeen spacetime, known as Bardeen–boson stars, which combine regularity with magnetic charge. Bardeen–boson stars arise from the coupling of a complex scalar field to the Bardeen geometry and can be regarded as a class of boson stars endowed with magnetic charge. Wang \emph{et al.} found that when the magnetic charge exceeds a critical value, the system approaches a zero-frequency limit, forming what is termed a frozen Bardeen–boson star~\cite{wang2023bardeen,huang2025orbits}. Such objects possess no event horizon but develop a “critical surface” at a specific radius, outside which the spacetime is nearly indistinguishable from that of an extremal Bardeen black hole. The photon geodesic structure reveals that a frozen Bardeen–boson star can generate a photon-ring-like region, where light remains trapped near the critical radius for an extended period, making its shadow and imaging features remarkably similar to those of a black hole. Furthermore, when a Dirac field is introduced, one can obtain a frozen Bardeen–Dirac star, which exhibits both stable and unstable photon-ring structures inside and outside the object~\cite{zhang2025bardeen}. These results indicate that the coupling between the scalar (or Dirac) field and the Bardeen electromagnetic field is mathematically consistent and physically natural, giving rise to horizonless yet extremely compact regular configurations. This provides a solid theoretical foundation for studying the interplay among spacetime regularity, horizon formation, and nonlinear electromagnetic effects in strong-gravity regimes. Therefore, the investigation of Bardeen–boson stars holds significant physical importance: by coupling the Bardeen spacetime of regular black holes with a complex scalar field, it reveals the possibility of constructing singularity-free and horizonless compact objects. The magnetic charge, serving as an additional degree of freedom, offers a new avenue to explore the coupling between NLED and scalar fields in strong gravitational fields, thereby deepening our understanding of their impact on the internal structure, stability, and external geometry of compact objects. From an observational perspective, studying the imaging properties of such boson stars provides theoretical guidance and diagnostic criteria for distinguishing different types of compact objects through high-resolution observations such as those from the EHT.

The structure of this paper is organized as follows. Section~\ref{sec2} presents the action including the complex scalar field and derives the numerical solutions for Bardeen-boson stars. Section~\ref{sec3} introduces the optical and geometrically thin accretion disk models, discussing the optical images of boson stars illuminated by these models as light sources. In Section~\ref{sec4}, we analyze the polarized images under synchrotron radiation conditions. Finally, Section~\ref{sec5} provides the conclusion and discussion. Throughout the paper, we use geometric units with $c = G = 1$, where $c$ is the vacuum speed of light and $G$ is the gravitational constant.

\section{The solutions and fitting of boson stars}\label{sec2}
This section briefly introduces the model of Einstein-Klein-Gordon theory coupled to  NLED, whose bulk action is
\begin{equation}
	S = \int \sqrt{-g}\, d^4x \left( \frac{R}{4} + L^{(1)} + L^{(2)} \right),\label{eq:act}
\end{equation}
where
\begin{align}
	L^{(1)} & = -\frac{3}{2\mathcal{S}}\left(\frac{\sqrt{2\mathcal{G}^{2}F}}{1+\sqrt{2\mathcal{G}^{2}F}}\right)^{\frac{5}{2}},\\
	L^{(2)} & = -\nabla_{a}\psi^{*}\nabla^{a}\psi-\mu^{2}\psi\psi^{*},
\end{align}
$R$ denotes the scalar curvature, the Lagrangian density $L^{(1)}$ is a function of $F$, and $F=\frac{1}{4}\mathcal{F}{ab}\mathcal{F}^{ab}$, where $\mathcal{F}{ab}=\partial_a\mathcal{A}_b-\partial_b\mathcal{A}_a$ and $\mathcal{A}$ is the electromagnetic field, $\psi$ is the complex scalar field. The constants $\mathcal{G}$, $\mathcal{S}$, and $\mu$ are three independent parameters, where $\mathcal{G}$ represents the magnetic charge, and $\mu$ denotes the mass of the scalar field. In this work, we set $\mu=1$. It should be noted that $L^{(1)}$ does not reduce to Maxwell electrodynamics. Consequently, the electromagnetic field associated with $L^{(1)}$ may differ from the one typically observed on Earth. In accordance with~\cite{zeng2025optical}, we assume that the Lagrangian density of the electromagnetic field (photons) observed on Earth retains the standard form of Maxwell electrodynamics. By varying the action~(\ref{eq:act}) with respect to the metric, the electromagnetic field, and the scalar field, one obtains the following equations of motion
\begin{align}
	R_{ab}-\frac{1}{2}g_{ab}R-2\left(T_{ab}^{(1)}+T_{ab}^{(2)}\right) & = 0,\label{eq:fe1}\\
	\nabla_{a}\left(\frac{\partial L^{(1)}}{\partial F}\mathcal{F}^{ab}\right) & = 0,\\
	\square\psi-\mu^{2}\psi & = 0,\label{eq:fe3}
\end{align}
where
\begin{align}
	T_{ab}^{(1)} & = -\frac{\partial L^{(1)}}{\partial F}\mathcal{F}_{ac}\mathcal{F}_{b}{}^{c}+g_{ab}L^{(1)},\\
	T_{ab}^{(2)} & = \partial_{a}\psi^{*}\partial_{b}\psi+\partial_{b}\psi^{*}\partial_{a}\psi
	-g_{ab}\left[\frac{1}{2}g^{ab}\left(\partial_{a}\psi^{*}\partial_{b}\psi+\partial_{b}\psi^{*}\partial_{a}\psi\right)+\mu^{2}\psi^{*}\psi\right].
\end{align}
Consider the static spherically symmetric solution
\begin{align}
	ds^2 &= g_{tt}dt^2 + g_{rr}dr^2 + g_{\theta\theta}d\theta^2 + g_{\varphi\varphi}d\phi^2 \nonumber\\
	&= -N(r)W^2(r)\,dt^2 + \frac{dr^2}{N(r)} + r^2\left(d\theta^2 + \sin^2\theta\,d\varphi^2\right), \label{eq:ssss}
\end{align}
where the functions $N(r)$ and $W(r)$ depend only on the radial coordinate $r$. The electromagnetic field and scalar field are taken as
\begin{equation}
	\mathcal{A}=\mathcal{G}\cos(\theta)\,d\varphi,\quad \psi=\phi(r)e^{-i\omega t}.\label{eq:esf}
\end{equation}
Substituting Eqs.~(\ref{eq:ssss}) and (\ref{eq:esf}) into Eqs.~(\ref{eq:fe1})--(\ref{eq:fe3}), the equations of motion are obtained as
\begin{align}
	\frac{W^{\prime}}{W}-2r\left(\phi^{\prime2}+\frac{\omega^{2}\phi^{2}}{N^{2}W^{2}}\right) & = 0,\\
	\phi^{\prime\prime}+\left(\frac{2}{r}+\frac{N^{\prime}}{N}+\frac{W^{\prime}}{W}\right)\phi^{\prime}+\left(\frac{\omega^{2}}{NW^{2}}-\mu^{2}\right)\frac{\phi}{N} & = 0,\\
	N^{\prime}+2\mu^{2}r\phi^{2}+\frac{2r\omega^{2}\phi^{2}}{NW^{2}}+2rN\phi^{\prime2}+\frac{N}{r}+\frac{3\mathcal{G}^{5}r}{\mathcal{S}\left(\mathcal{G}^{2}+r^{2}\right)^{5/2}}-\frac{1}{r} & = 0.
\end{align}
The above system of differential equations can be solved numerically by using the shooting method. For this purpose, appropriate boundary conditions must be imposed: at spatial infinity, the solution should satisfy the asymptotic flatness of the Schwarzschild spacetime; at $r\to0$, the solution must remain nonsingular. Thus,
\begin{equation}
	N(0)=1,\quad W(0)=W_0,\quad N(\infty)=1-\frac{2M}{r},\quad W(\infty)=1,
\end{equation}
where $M$ denotes the Arnowitt-Deser-Misner (ADM) mass and $W_0$ is a constant, both of which can be obtained by solving the system. In addition, for the complex scalar field, it is required that
\begin{equation}
	\phi(\infty)=0,\quad \left.\frac{d\phi(r)}{dr}\right|_{r=0}=0.
\end{equation}
The solution for the boson star can be characterized by the ADM mass $M$ and the radius $R$, which are defined via the mass function $\mathcal{M}=\frac{r}{2}\left[1-N(r)\right]$ as
\begin{equation}
	M=\mathcal{M}(r\to\infty),\quad R=0.98M.
\end{equation}
Since the mass distribution of the boson star is highly concentrated, the results for the radius obtained by different definitions differ only slightly.

Before discussing the imaging properties of boson stars, it is instructive to study the geodesic structure of the spacetime. For a static and spherically symmetric spacetime~(\ref{eq:ssss}), the Lagrangian of a test particle is given by
\begin{equation}
	\tilde{\mathcal{L}}=-\frac{1}{2}g_{\mu\nu}\dot{x}^\mu\dot{x}^\nu=\frac{\eta}{2},
\end{equation}
where $\eta=0$ corresponds to photons and $\eta=1$ to timelike particles. Owing to the spherical symmetry of the system, we restrict the motion to the equatorial plane ($\theta=\pi/2$), where the Lagrangian reduces to
\begin{equation}
	\tilde{\mathcal{L}}=-\frac{1}{2}\left[g_{tt}(r)\dot{t}^{2}+g_{rr}(r)\dot{r}^{2}+g_{\varphi\varphi}(r)\dot{\varphi}^{2}\right],
\end{equation}
with the overdot denoting derivatives with respect to the affine parameter. The spacetime admits a timelike and a spacelike Killing vector, which lead to two conserved quantities
\begin{equation}
	\mathcal{E}=\frac{\partial\tilde{\mathcal{L}}}{\partial\dot{t}}=-g_{tt}\dot{t},\quad
	\mathcal{L}=-\frac{\partial\tilde{\mathcal{L}}}{\partial\dot{\varphi}}=g_{\varphi\varphi}\dot{\varphi}=r^2\dot{\varphi},\label{eq:el}
\end{equation}
representing, respectively, the energy and angular momentum per unit mass. These relations yield the geodesic equations
\begin{align}
	\dot{t} & =-\frac{\mathcal{E}}{g_{tt}} ,\\
	\dot{r} & = \pm \left[-\frac{1}{g_{rr}}\left(\eta+\frac{1}{g_{tt}}\mathcal{E}^2+\frac{1}{g_{\varphi\varphi}}\mathcal{L}^2\right)\right]^{\frac{1}{2}},\\
	\dot{\theta}& = 0,\\
	\dot{\varphi} & = \frac{\mathcal{L}}{g_{\varphi\varphi}}.
\end{align}
Based on these relations, the effective potential can be defined as
\begin{align}
	\mathcal{V}(r) & =\dot{r}^2 \\& = -\frac{1}{g_{rr}}\left(\eta+\frac{1}{g_{tt}}\mathcal{E}^2+\frac{1}{g_{\varphi\varphi}}\mathcal{L}^2\right).\label{eq:veff}
\end{align}
For photons ($\eta=0$), this becomes
\begin{equation}
	\mathcal{V}_{photon}(r) = -\frac{1}{g_{rr}}\left(\frac{1}{g_{tt}}\mathcal{E}^2+\frac{1}{g_{\varphi\varphi}}\mathcal{L}^2\right).\label{eq:veffp}
\end{equation}
In the case of black holes, we focus on the unstable null circular orbits near the event horizon, known as photon rings~\cite{synge1966escape,luminet1979image,dewitt1973black,cunningham1972optical}. As an intrinsic property of the black hole, the position of the photon ring depends only on the spacetime parameters and remains unchanged with the observation angle $\theta_o$. A similar concept applies to boson stars, where the photon ring radius can be determined from the effective potential and its derivative
\begin{equation}
	\mathcal{V}_{photon}(r) = 0,\quad \partial_r \mathcal{V}_{photon}(r)=0.
\end{equation}
The stability of the photon ring is then characterized by the second derivative
\begin{align}
	\partial^2_r \mathcal{V}_{photon}(r)\begin{cases}<0,&\text{unstable,}\\=0,&\text{marginally stable,}\\>0,&\text{stable.}\end{cases}
\end{align}

In numerical computations, due to the presence of numerical infinity, it is inconvenient to directly use the numerical metric for subsequent calculations. Therefore, in this paper, a fitting method is adopted to obtain an analytical form of the metric, and the fitting functions are expressed as
\begin{align}
	g_{tt} & = -\exp\left\{a_{7}\left[\exp\left(-\frac{1+a_{1}r+a_{2}r^{2}}{a_{3}+a_{4}r+a_{5}r^{2}+a_{6}r^{3}}\right)-1\right]\right\},\\
	g_{rr} & = \exp\left\{b_{7}\left[\exp\left(-\frac{1+b_{1}r+b_{2}r^{2}}{b_{3}+b_{4}r+b_{5}r^{2}+b_{6}r^{3}}\right)-1\right]\right\}.
\end{align}
These coefficients must satisfy the boundary conditions of the equations of motion and are thus not completely independent. In this paper, we systematically study the effects of the initial scalar field $\phi_0$ and the magnetic charge $\mathcal{G}$ by selecting different values of $\phi_0$ and $\mathcal{G}$, considering eight sets of boson stars in total. When the constants $\mathcal{S}=0.8$ and $\mathcal{G}=0.35$, $\phi_0=(0.3,0.35,0.55,0.6)$ correspond to BS1, BS2, BS3, and BS4, respectively. When the constants $\mathcal{S}=0.2$ and $\phi_0=0.6$, $\mathcal{G}=(0.01,0.07,0.1,0.15)$ correspond to BS5, BS6, BS7, and BS8, respectively.

\begin{figure}[!h]
	\centering
	\subfigure[]{\includegraphics[scale=0.5]{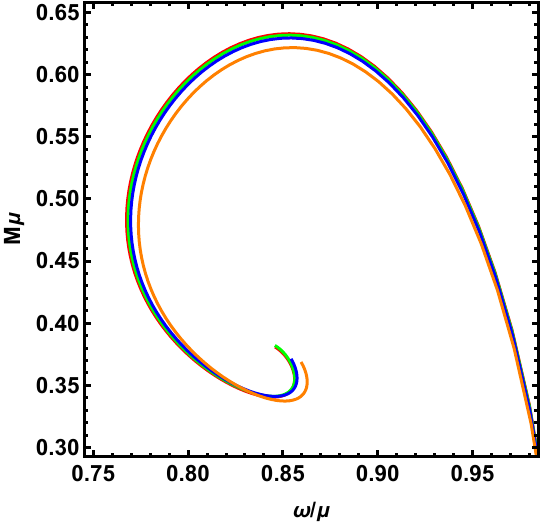}} 
	\subfigure[]{\includegraphics[scale=0.5]{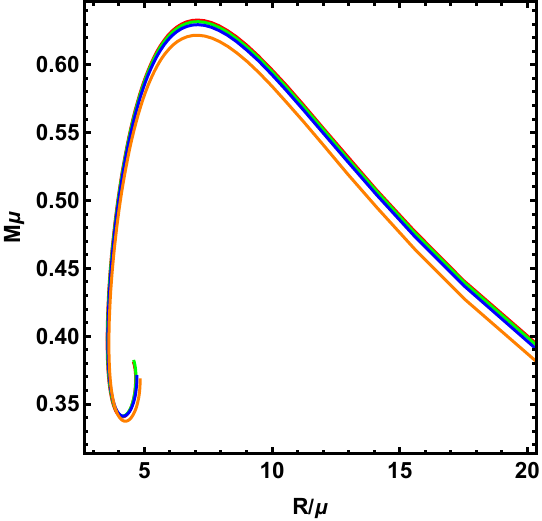}} 
	\subfigure[]{\includegraphics[scale=0.51]{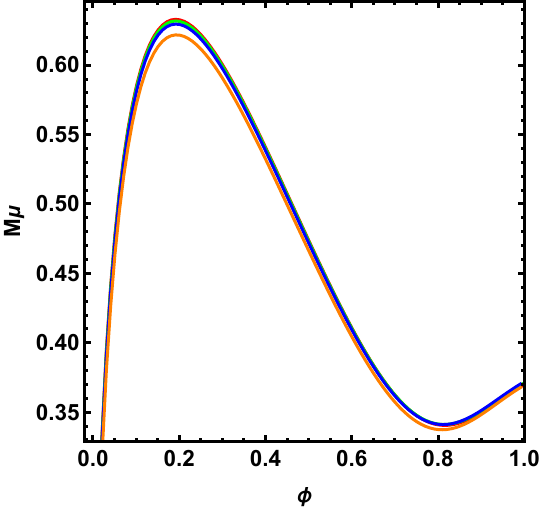}} 
	
	\caption{Relationship between the mass and the frequency $\omega$, radius $R$, and scalar field $\phi$. The red, green, blue, and orange lines correspond to $\mathcal{G}=0.01, 0.07, 0.1,$ and $0.15$, respectively. The fixed parameters are $\mathcal{S}=0.2$ and $\phi_0=0.6$.}
	\label{fig0}
\end{figure}

\begin{figure}[!h]
	\centering 
	\subfigure[]{\includegraphics[scale=0.52]{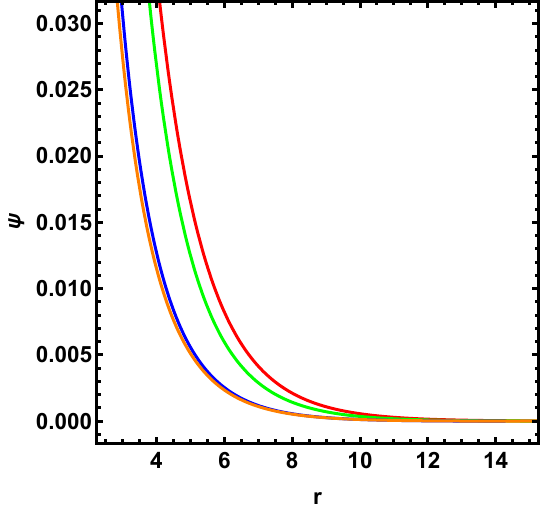}}
	\subfigure[]{\includegraphics[scale=0.5]{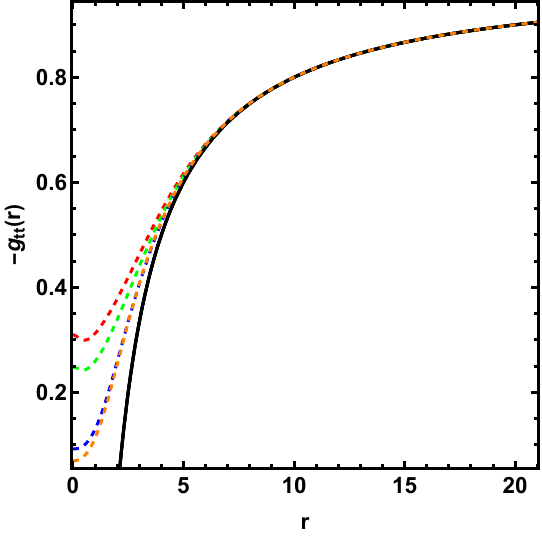}}
	\subfigure[]{\includegraphics[scale=0.5]{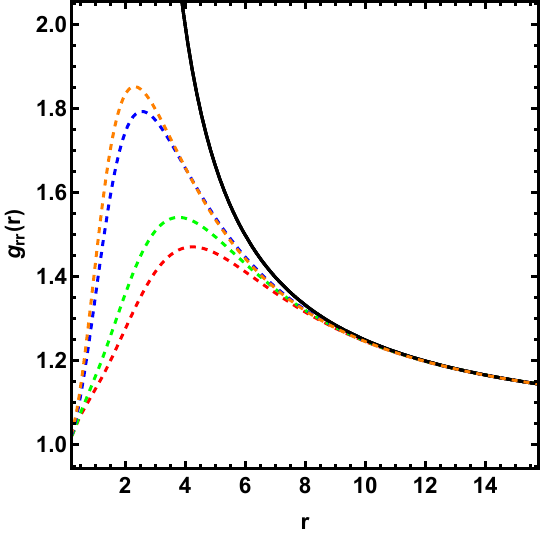}}
	
	\caption{Scalar field and numerical metric of BS1--BS4. The red, green, blue, and orange lines correspond to $\phi_0=0.3,0.35,0.55,0.6$, respectively. The fixed parameters are $\mathcal{S}=0.8,\mathcal{G}=0.35$.}
	\label{fig1}
\end{figure}

\begin{figure}[!h]
	\centering
	\subfigure[]{\includegraphics[scale=0.52]{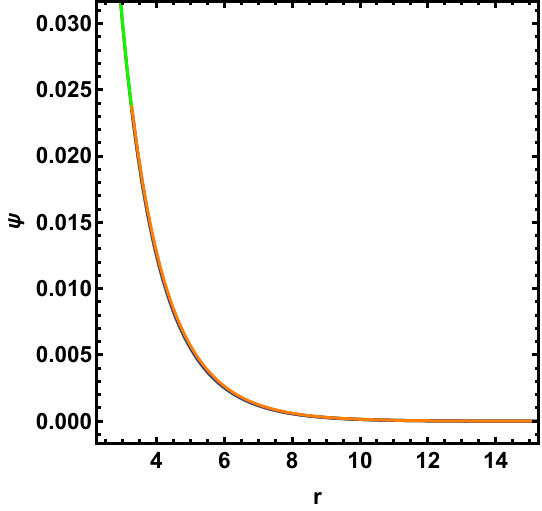}}
	\subfigure[]{\includegraphics[scale=0.5]{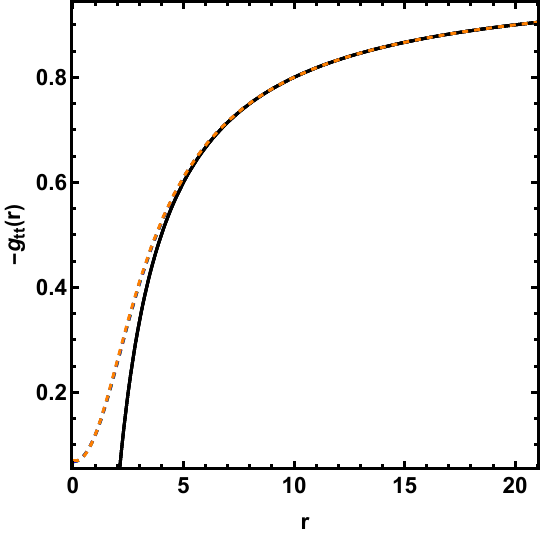}}
	\subfigure[]{\includegraphics[scale=0.5]{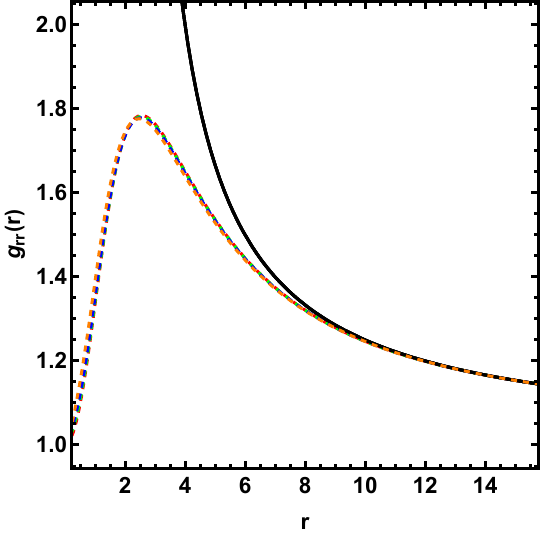}}
	
	\caption{Scalar field and numerical metric of BS5--BS8. The red, green, blue, and orange lines correspond to $\mathcal{G}=0.01,0.07,0.1,0.15$, respectively. The fixed parameters are $\mathcal{S}=0.2,\phi_0=0.6$.}
	\label{fig2}
\end{figure}

\begin{figure}[!h]
	\centering 
	\subfigure[]{\includegraphics[scale=0.515]{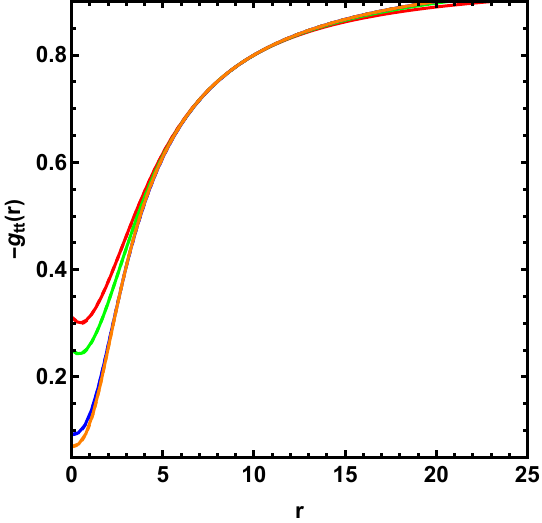}}
	\subfigure[]{\includegraphics[scale=0.5]{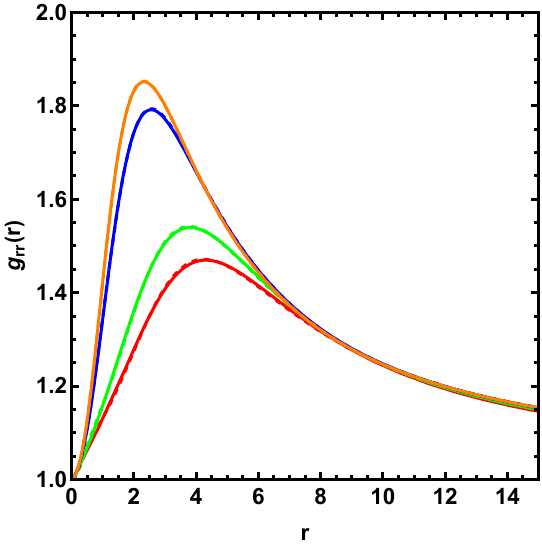}}
	
	\caption{Fitting results for BS1--BS4. The solid lines represent the fitting functions, and the dashed lines represent the numerical metrics. The red, green, blue, and orange lines correspond to $\phi_0=0.3,0.35,0.55,0.6$, respectively. The fixed parameters are $\mathcal{S}=0.8,\mathcal{G}=0.35$.}
	\label{fig3}
\end{figure}

\begin{figure}[!h]
	\centering 
	\subfigure[]{\includegraphics[scale=0.515]{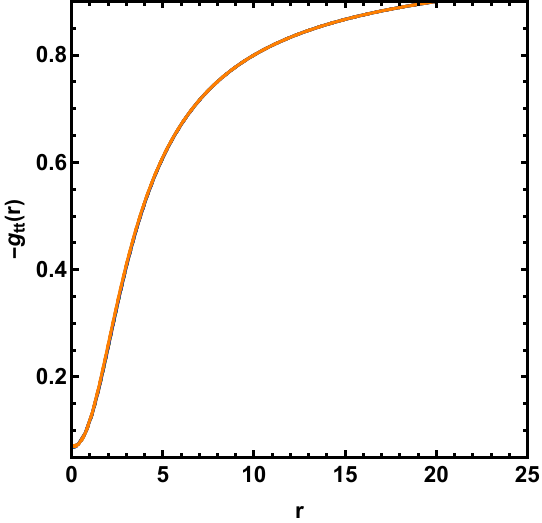}}
	\subfigure[]{\includegraphics[scale=0.5]{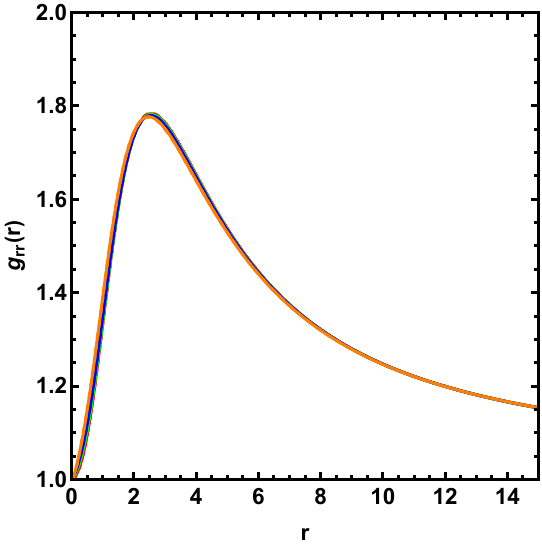}}
	
	\caption{Fitting results for BS5--BS8. The solid lines represent the fitting functions, and the dashed lines represent the numerical metrics. The red, green, blue, and orange lines correspond to $\mathcal{G}=0.01,0.07,0.1,0.15$, respectively. The fixed parameters are $\mathcal{S}=0.2,\phi_0=0.6$.}
	\label{fig4}
\end{figure}

Fig.~\ref{fig0} illustrates the relationships between the mass $M$ of the boson star and $(\omega, R, \phi)$ for different values of $\mathcal{G}$. It can be seen that as $\omega$ and $R$ decrease, $M$ first increases and then decreases, exhibiting a spiral behavior. When $\phi$ increases, $M$ initially grows, then decreases, and rises again. For fixed values of $\omega$, $R$, and $\phi$, an increase in $\mathcal{G}$ leads to a reduction in $M$. Fig.~\ref{fig1} and~\ref{fig2} show the variation of the scalar field $\phi$ and the numerical metric of BS1--BS8 with respect to the radial coordinate $r$. In the figures, the black solid line represents the metric components of the Schwarzschild solution with unit mass. As $r$ increases, $\phi$ rapidly approaches zero, and the numerical metric gradually converges to the Schwarzschild metric, demonstrating the asymptotic flatness of the spacetime. Tables~\ref{tab1}--\ref{tab4} present the parameter estimates of the fitting functions and the masses of the boson stars. The results show that the boson star mass $M$ decreases with increasing $\phi_0$ and $\mathcal{G}$. Fig.~\ref{fig3} and~\ref{fig4} display the comparison between the numerical metric and the fitting functions as functions of $r$, where the solid lines denote the fitting functions and the dashed lines denote the numerical metric. It is evident that the difference between the two is negligible, indicating that it is reasonable to use the fitting functions in place of the numerical metric for subsequent calculations.

\begin{table}[ht]
	\centering
	\caption{The Bardeen-boson stars for $\mathcal{S}=0.8,\mathcal{G}=0.35$.}
	\label{tab1}
	\begin{tabular}{c|c|c|c|c|c|c|c|c|c}
		\hline
		Type & $\phi_0$ & $M$ & $a_1$ & $a_2$ & $a_3$ & $a_4$ & $a_5$ & $a_6$ & $a_7$ \\ \hline
		BS1 & 0.3  & 0.576572 & 0.410  & 0.096  & -0.096  & -0.037  & -0.013  & 0.000  & 0.000 \\ \hline
		BS2 & 0.35 & 0.550220 & 0.342  & 0.121  & -0.222  & -0.065  & -0.051  & -0.002 & -0.016 \\ \hline
		BS3 & 0.55 & 0.423811 & 0.772  & 1.002  & -0.617  & -0.465  & -0.767  & -0.693 & -0.586 \\ \hline
		BS4 & 0.6  & 0.394075 & 0.973  & 1.199  & -20538.1 & -20272.8 & -36318.8 & -83020.9 & -54592.0 \\ \hline
	\end{tabular}
\end{table}

\begin{table}[ht]
	\centering
	\caption{The Bardeen-boson stars for $\mathcal{S}=0.8,\mathcal{G}=0.35$.}
	\label{tab2}
	\begin{tabular}{c|c|c|c|c|c|c|c|c|c}
		\hline
		Type & $\phi_0$ & $M$ & $b_1$ & $b_2$ & $b_3$ & $b_4$ & $b_5$ & $b_6$ & $b_7$ \\ \hline
		BS1 & 0.3  & 0.576572 & -30.715  & -72.865  & 51.991   & 281.036  & -73.258  & 43.386  & -0.531 \\ \hline
		BS2 & 0.35 & 0.550220 & -43.727  & -30.037  & 219.403  & 100.956  & -15.235  & 38.952  & -1.196 \\ \hline
		BS3 & 0.55 & 0.423811 & -48.222  & -16.671  & 4.185    & 7.584    & 7.803    & 0.592   & -0.025 \\ \hline
		BS4 & 0.6  & 0.394075 & -53.674  & -20.747  & 2.615    & 8.073    & 7.961    & 0.465   & -0.013 \\ \hline
	\end{tabular}
\end{table}

\begin{table}[ht]
	\centering
	\caption{The Bardeen-boson stars for $\mathcal{S}=0.2,\phi_0=0.6$.}
	\label{tab3}
	\begin{tabular}{c|c|c|c|c|c|c|c|c|c}
		\hline
		Type & $\mathcal{G}$   & $M$      & $a_1$   & $a_2$   & $a_3$   & $a_4$   & $a_5$   & $a_6$   & $a_7$    \\ \hline
		BS5  & 0.01  & 0.410811 & 0.766   & 0.916   & -0.436  & -0.341  & -0.609  & -0.337  & -0.302   \\ \hline
		BS6  & 0.07  & 0.410463 & 0.766   & 0.859   & -0.409  & -0.315  & -0.562  & -0.267  & -0.255   \\ \hline
		BS7  & 0.1  & 0.409360 & 0.818   & 0.799   & -0.384  & -0.307  & -0.530  & -0.209  & -0.214   \\ \hline
		BS8  & 0.15  & 0.404415 & 1.070   & 0.800   & -0.378  & -0.375  & -0.591  & -0.201  & -0.203   \\ \hline
	\end{tabular}
\end{table}

\begin{table}[ht]
	\centering
	\caption{The Bardeen-boson stars for $\mathcal{S}=0.2,\phi_0=0.6$.}
	\label{tab4}
	\begin{tabular}{c|c|c|c|c|c|c|c|c|c}
		\hline
		Type & $\mathcal{G}$   & $M$      & $b_1$     & $b_2$     & $b_3$   & $b_4$   & $b_5$   & $b_6$   & $b_7$      \\ \hline
		BS5  & 0.01  & 0.410811 & -47.921   & -12.693   & 2.495   & 6.907   & 5.230   & 0.198   & -0.011     \\ \hline
		BS6  & 0.07  & 0.410463 & -79.014   & -31.747   & 6.175   & 11.899  & 13.241  & 0.956   & -0.020     \\ \hline
		BS7  & 0.1  & 0.409360 & -163.586  & -84.319   & 17.350  & 26.462  & 36.977  & 4.039   & -0.033     \\ \hline
		BS8  & 0.15  & 0.404415 & -71.169   & -69.057   & 13.020  & 18.429  & 31.108  & 8.880   & -0.095     \\ \hline
	\end{tabular}
\end{table}

\section{Images of boson stars under the thin disk}\label{sec3}
In this section, we discuss the optical images of Bardeen-boson stars under both optical and geometrically thin accretion disks. The accretion disk located in the equatorial plane serves as the light source, where the plasma on the disk moves in circular timelike orbits. In this case, the effective potential~(\ref{eq:veff}) takes $\eta=1$, and the position of the innermost stable circular orbit (ISCO) of the boson star, denoted by $r_I$, can be determined from the effective potential as
	\begin{equation}
		\mathcal{V}_{particle}(r)=0|_{r=r_I},\quad\partial_r\mathcal{V}_{particle}(r)=0|_{r=r_I},\quad\partial_r^2\mathcal{V}_{particle}(r)=0|_{r=r_I}.
	\end{equation}
Inside the ISCO, it is assumed that there is no accreting matter, i.e., no light source. Near the outer side of the ISCO, the accretion flow moves along circular timelike orbits, with four-velocity
\begin{equation}
	u^\mu=\left(\sqrt{\frac{2g_{tt}^2}{r\partial_rg_{tt}-2g_{tt}}},0,0,\sqrt{\frac{r^3\partial_rg_{tt}}{2g_{tt}-r\partial_rg_{tt}}}\right),
\end{equation}
On the other hand, the motion of photons is governed by the null geodesic equations, which are numerically solved in this paper using the Hamiltonian formalism. For photons emitted from the source and arriving at the observer, their four-momentum can be described in a locally static coordinate system. This frame is located at the observer and is called as the zero angular momentum observer (ZAMO) tetrad, which takes the form
\begin{equation}
	e^{\nu}{}_{(\mu)}=\begin{pmatrix}\sqrt{-\frac1{g_{tt}}}&0&0&0\\0&-\sqrt{\frac1{g_{rr}}}&0&0\\0&0&\sqrt{\frac1{g_{\theta\theta}}}&0\\0&0&0&-\sqrt{\frac1{g_{\varphi\varphi}}}\end{pmatrix}.\label{eq:tetrad}
\end{equation}
In the ZAMO tetrad, the four-momentum of the photon can be expressed as $p_{(\mu)}=p_{\nu}e^{\nu}{}_{(\mu)}$, where $p_{(\mu)}$ and $p_\nu$ denote the four-momentum components in the ZAMO and Boyer-Lindquist coordinates, respectively. At the observer, the celestial coordinates $(\alpha,\beta)$ are introduced, with the celestial radius given by the magnitude of the photon’s spatial momentum. The position of the photon on the screen can be obtained by stereographic projection techniques. According to Ref.~\cite{hu2021qed}, the celestial coordinates can be written as
\begin{equation}
	\cos\alpha=\frac{p^{(r)}}{p^{(t)}},\quad\tan\beta=\frac{p^{(\varphi)}}{p^{(\theta)}},
\end{equation}
A Cartesian coordinate system $(x,y)$ can also be established on the projection screen, which is related to the celestial coordinates by
\begin{equation}
	x=-2\tan\frac\alpha2\sin\beta,\quad y=-2\tan\frac\alpha2\cos\beta.\label{eq:cvs}
\end{equation}
The field of view $\gamma_{\mathrm{fov}}$ determines the visible range of the camera. For convenience, in the Cartesian coordinates, the ranges along the $x$ and $y$ directions are taken as $\gamma_{\mathrm{fov}}/2$, defining a square screen with side length
\begin{equation}
	L=2|\overrightarrow{OP}|\tan\frac{\gamma_{\mathrm{fov}}}{2},
\end{equation}
where $\overrightarrow{OP}$ denotes the three-dimensional momentum of the photon. The imaging plane is divided into $n\times n$ pixels, with each pixel having a side length
\begin{equation}
	l=\frac{L}{n}=\frac{2}{n}|\overrightarrow{OP}|\tan\frac{\gamma_{\mathrm{fov}}}{2}.
\end{equation}
The center of each pixel is labeled by coordinates $(i,j)$, with the bottom-left pixel denoted by $(1,1)$ and the top-right by $(n,n)$, where $i$ and $j$ run from $1$ to $n$, respectively. The relationship between the Cartesian coordinates $(x,y)$ and the pixel coordinates $(i,j)$ is
\begin{equation}
	x=l\left(i-\frac{n+1}{2}\right),\quad y=l\left(j-\frac{n+1}{2}\right).\label{eq:cvp}
\end{equation}
Comparing Eqs.~(\ref{eq:cvs}) and (\ref{eq:cvp}), the correspondence between the pixel coordinates $(i,j)$ and the celestial coordinates $(\alpha,\beta)$ is obtained as
\begin{align}
	\tan\frac{\alpha}{2}&=\frac{1}{n}\tan\left(\frac{\gamma_{\mathrm{fov}}}{2}\right)\sqrt{\left(i-\frac{n+1}{2}\right)^2+\left(j-\frac{n+1}{2}\right)^2},\\
	\tan\beta&=\frac{2j-(n+1)}{2i-(n+1)}.
\end{align}

To study the optical images of Bardeen-boson stars, it is necessary to choose an appropriate source model. Based on current astronomical observations, geometrically and optically thin accretion disks are typically used as sources for black hole imaging~\cite{gralla2019black}. Since both black holes and boson stars are compact objects, the same source model is adopted in this work. When light interacts with the accretion disk, the emission and absorption of photons cause changes in the radiation intensity. In this case, the radiative transfer equation for unpolarized radiation is given by
\begin{equation}
	\frac{d}{d\lambda}\left(\frac{s_\nu}{\nu^3}\right) = \frac{e_\nu - a_\nu s_\nu}{\nu^2},
\end{equation}
where $\lambda$ is the affine parameter of the photon, and $s_{\nu}$, $e_{\nu}$, and $a_{\nu}$ denote the specific intensity, emissivity, and absorption coefficient at frequency $\nu$, respectively. Under the thin accretion disk approximation, only the instantaneous emission and absorption of photons on the equatorial plane need to be considered, so that $e_{\nu}=a_{\nu}=0$ in regions off the equatorial plane, and $s_{\nu}/\nu^3$ is conserved along the null geodesic. According to Ref.~\cite{zhang2024observational}, a light ray may interact with the accretion disk multiple times, with each interaction contributing to the observed brightness. The total intensity observed on the screen is the sum of contributions from all such interactions, that is,
\begin{equation}
	s_o = \sum_{n=1}^{N} f_n e_n (\chi_n)^3,
\end{equation}
where $n=1,\ldots,N$ denotes the number of times the light ray crosses the equatorial plane, and $f_n$, $e_n$, and $\chi_n = \nu_o / \nu_n$ are the fudge factor, emissivity, and redshift factor, respectively. For $n=1$, the direct image is formed; for $n=2$, the lensed image is formed; and for $n\geqslant 2$, higher-order images are formed.

It can be seen from the above equation that the intensity $s_o$ is jointly determined by $f_n$, $e_n$, and $\chi_n$. Their values are briefly discussed below. The fudge factor $f_n$ is used to describe the absorption properties of the accretion disk and needs to be determined according to the specific accretion disk model. Previous studies have shown that $f_n$ mainly affects the intensity of higher-order images and has a limited impact on the overall optical image. Therefore, $f_n$ is set to $1$ in this work~\cite{he2025optical}. For different accretion disk models, various forms can be chosen for the emissivity $e_n$. To match astronomical observations (such as the imaging results of M87$^{\star}$ and Sgr A$^{\star}$), $e_n$ is typically assumed to be a second-order polynomial in log-space. For example, in the Gralla-Lupsasca-Marrone model~\cite{gralla2020shape},
\begin{equation}
	e_n = \frac{\exp\left[-\frac{1}{2}\left(c_1\operatorname{arcsinh}\left(\frac{r-c_2}{c_3}\right)\right)^2\right]}{\sqrt{(r-c_2)^2+(c_3)^2}},
\end{equation}
where $c_1$, $c_2$, and $c_3$ correspond to the rate of increase, radial translation, and dilation of the profile, respectively. This model is consistent with the accretion disk structure simulated by general relativistic magnetohydrodynamics and is therefore widely used~\cite{vincent2022images}. These parameters jointly control the spatial distribution of the emitted photons and can, in principle, be adjusted to match the numerical simulation results with actual observations. In this paper, we choose $c_1=0$, $c_2=6M$, and $c_3=M$, where $M$ is the mass of the boson star.

The redshift factor $\chi_n$ has a significant impact on the intensity $s_o$~\cite{luis2022shadows,rosa2023imaging}. $\chi_n$ includes contributions from both Doppler and gravitational redshift. Among these, Doppler redshift plays a dominant role in explaining the brightness distribution of the image, while gravitational redshift reflects the effect of the strong gravitational field on the propagation path of light. Assuming the accreting matter is electrically neutral plasma moving along timelike geodesics, similar to Eq.~(\ref{eq:el}), there are two conserved quantities along this geodesic. Outside the ISCO, the angular velocity of the fluid is given by
\begin{equation}
	\tilde{\omega}_n = \left.\frac{u^\varphi}{u^t}\right|_{r=r_n},
\end{equation}
where $r_n$ is the radial coordinate at which the light ray crosses the equatorial plane for the $n$th time. The redshift factor $\chi_n$ can be rewritten as
\begin{equation}
	\chi_n = -\frac{k}{K(1-\mathcal{I}\tilde{\omega}_n)},
\end{equation}
where $\mathcal{I}={|\mathcal{L}|}/{\mathcal{E}}$ is the photon impact parameter (see Eq.~(\ref{eq:el})), and $k$ is the ratio of the observed energy on the screen to $\mathcal{E}$. For asymptotically flat spacetimes, when the observer is located at infinity, $k=1$. $K$ is defined as
\begin{equation}
	K = \left.\sqrt{\frac{1}{g_{tt}+g_{\varphi\varphi}\tilde{\omega }_{n}^{2}}}\right|_{r=r_n}.
\end{equation}
With the above analysis, we can quantitatively calculate the intensity at each pixel.

\begin{figure}[!h]
	\centering 
	\subfigure[$\phi_0=0.3,\theta_o=0^\circ$]{\includegraphics[scale=0.35]{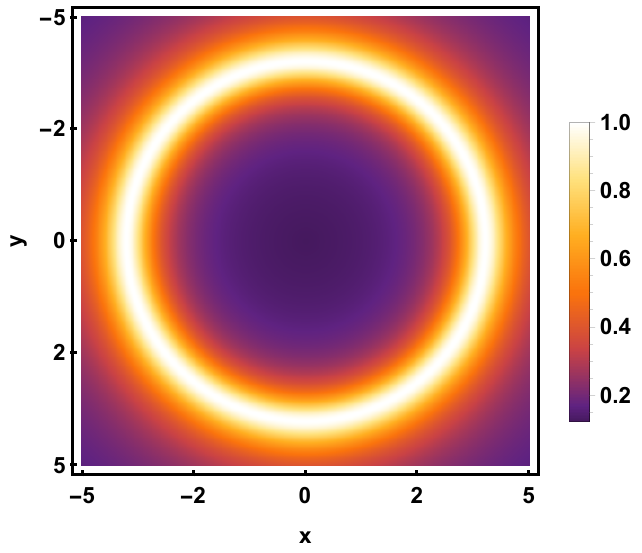}}
	\subfigure[$\phi_0=0.3,\theta_o=30^\circ$]{\includegraphics[scale=0.35]{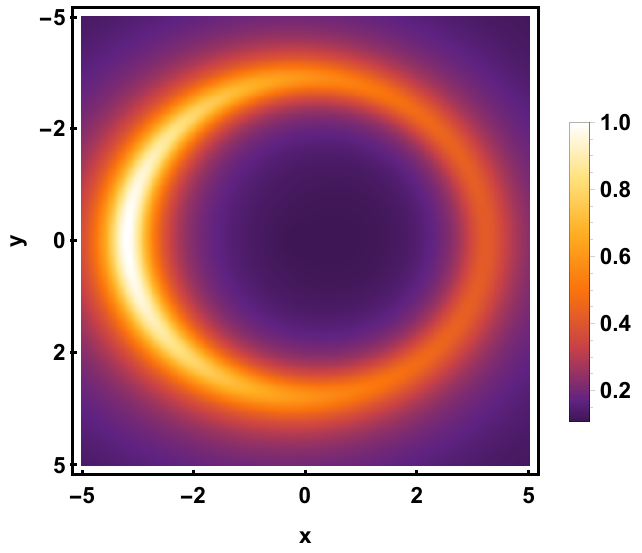}}
	\subfigure[$\phi_0=0.3,\theta_o=60^\circ$]{\includegraphics[scale=0.35]{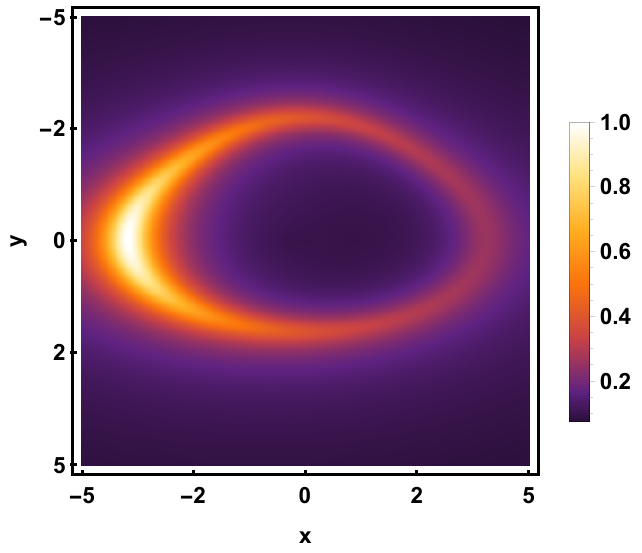}}
	\subfigure[$\phi_0=0.3,\theta_o=80^\circ$]{\includegraphics[scale=0.35]{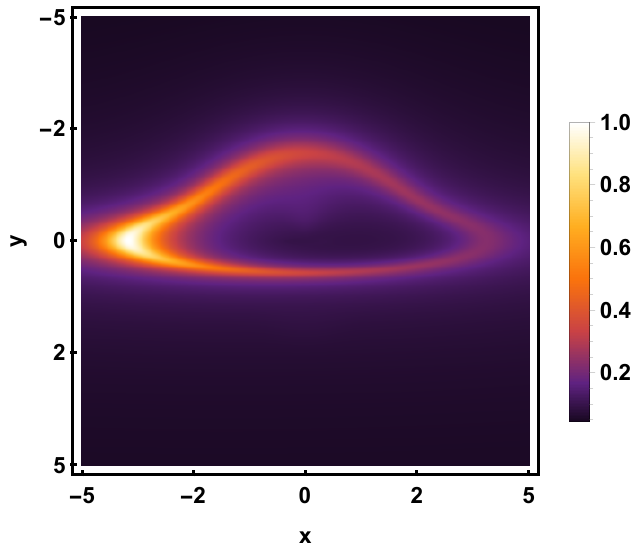}}
	
	\subfigure[$\phi_0=0.35,\theta_o=0^\circ$]{\includegraphics[scale=0.35]{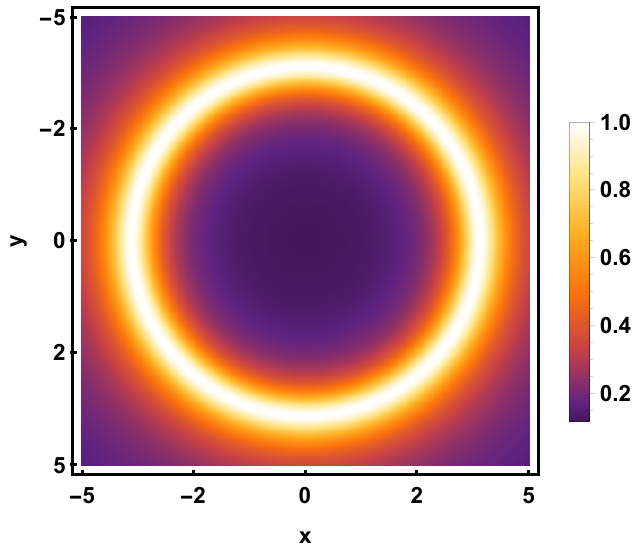}}
	\subfigure[$\phi_0=0.35,\theta_o=30^\circ$]{\includegraphics[scale=0.35]{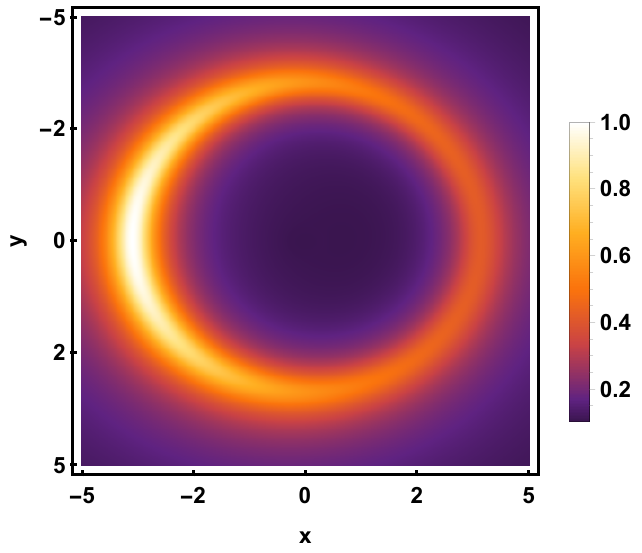}}
	\subfigure[$\phi_0=0.35,\theta_o=60^\circ$]{\includegraphics[scale=0.35]{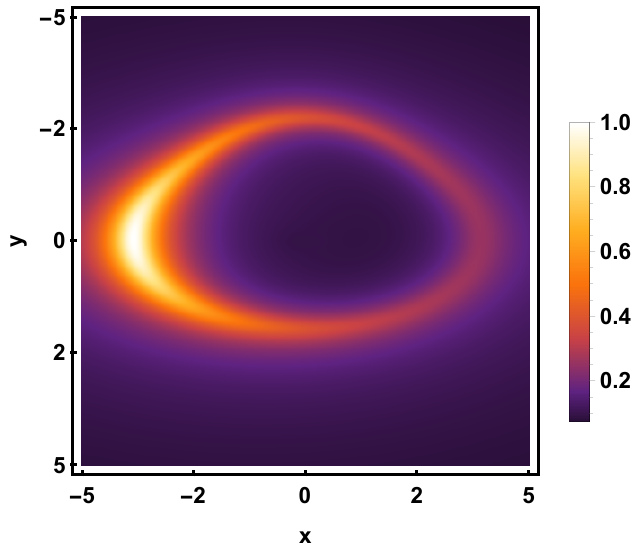}}
	\subfigure[$\phi_0=0.35,\theta_o=80^\circ$]{\includegraphics[scale=0.35]{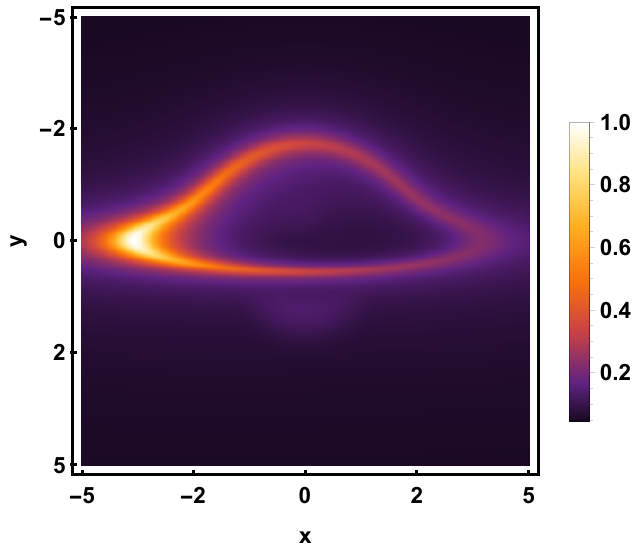}}
	
	\subfigure[$\phi_0=0.55,\theta_o=0^\circ$]{\includegraphics[scale=0.35]{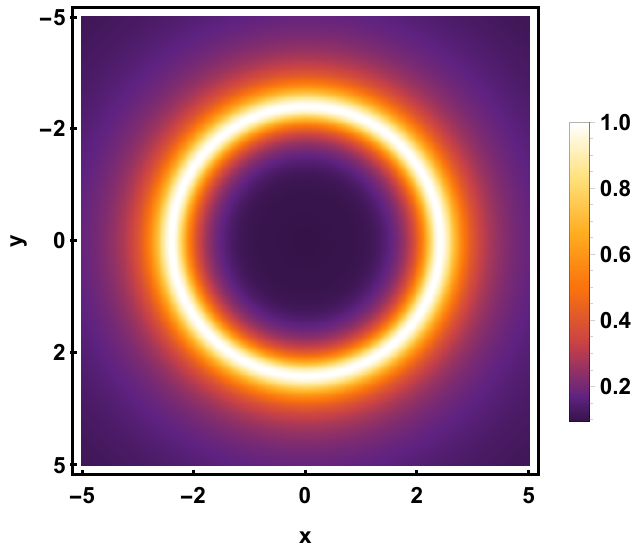}}
	\subfigure[$\phi_0=0.55,\theta_o=30^\circ$]{\includegraphics[scale=0.35]{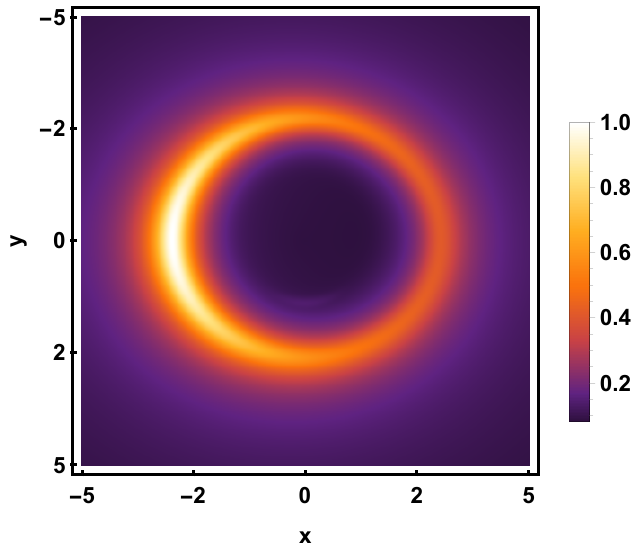}}
	\subfigure[$\phi_0=0.55,\theta_o=60^\circ$]{\includegraphics[scale=0.35]{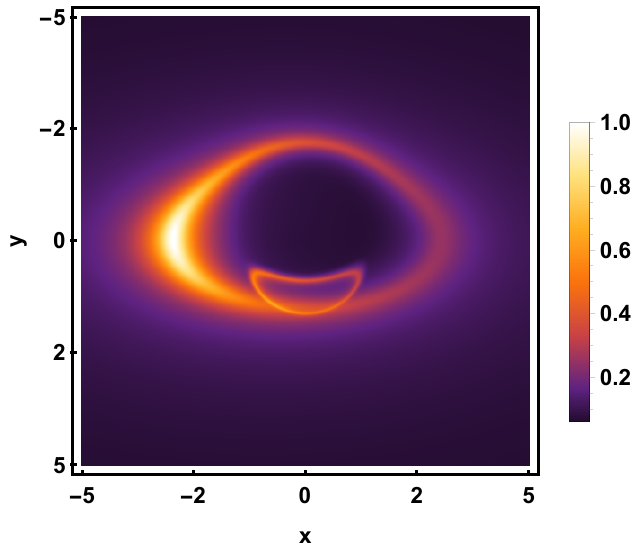}}
	\subfigure[$\phi_0=0.55,\theta_o=80^\circ$]{\includegraphics[scale=0.35]{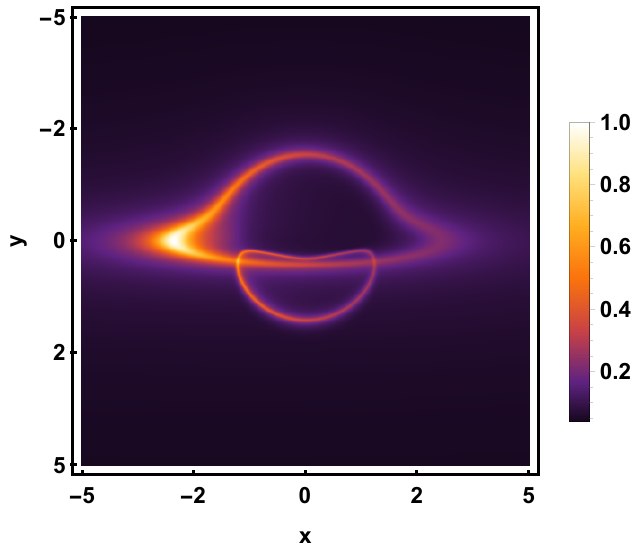}}
	
	\subfigure[$\phi_0=0.6,\theta_o=0^\circ$]{\includegraphics[scale=0.35]{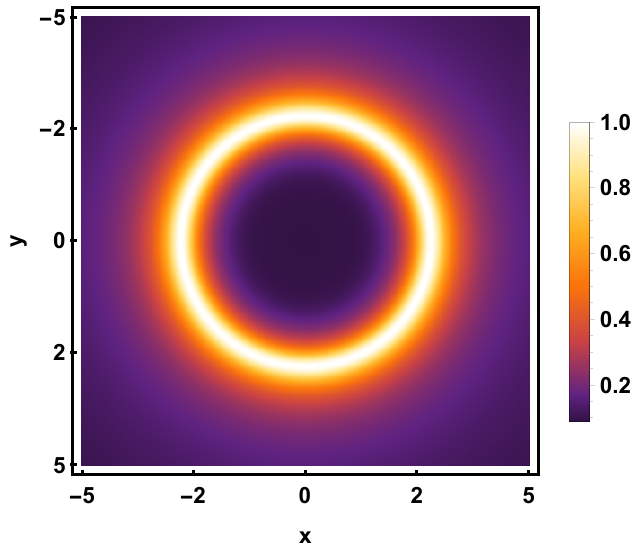}}
	\subfigure[$\phi_0=0.6,\theta_o=30^\circ$]{\includegraphics[scale=0.35]{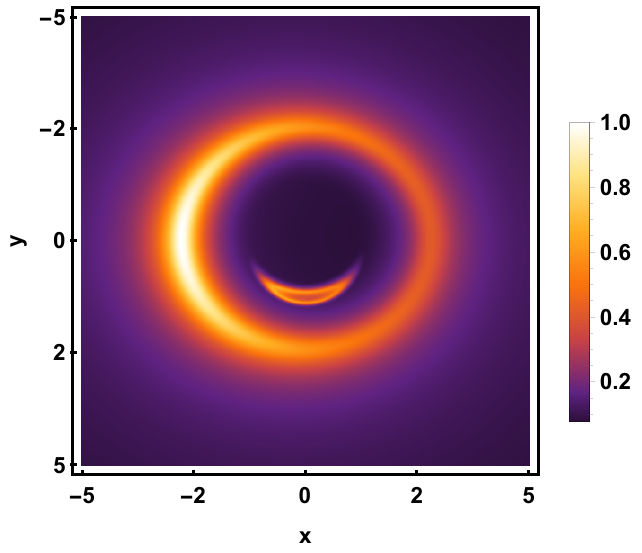}}
	\subfigure[$\phi_0=0.6,\theta_o=60^\circ$]{\includegraphics[scale=0.35]{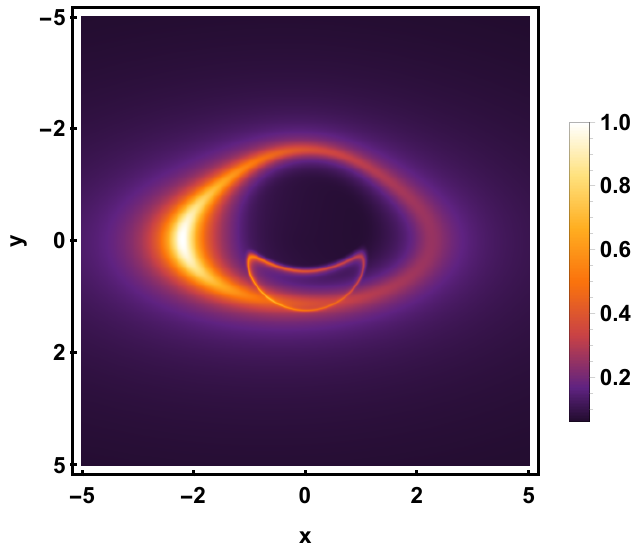}}
	\subfigure[$\phi_0=0.6,\theta_o=80^\circ$]{\includegraphics[scale=0.35]{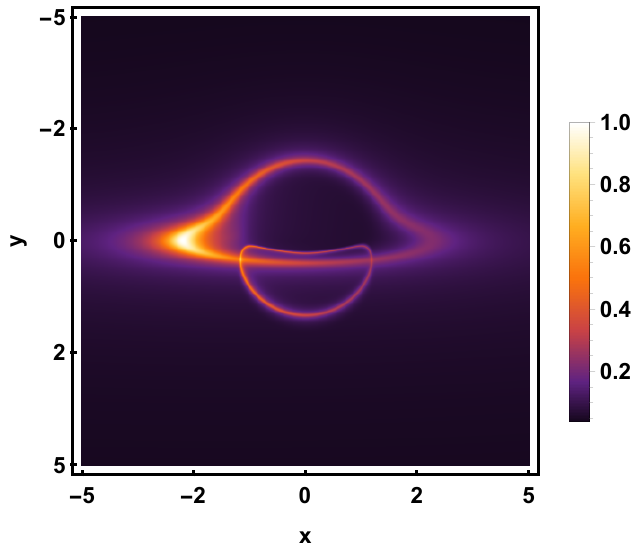}}
	
	\caption{Optical images of boson stars under the thin accretion disk. Each row from top to bottom corresponds to initial scalar fields $\phi_0=0.3,0.35,0.55,0.6$, and each column from left to right corresponds to observer inclination angles $\theta_o=0^\circ, 30^\circ, 60^\circ, 80^\circ$. The fixed parameters are $\mathcal{S}=0.8,\mathcal{G}=0.35,\gamma_{\mathrm{fov}}=3^\circ$.}
	\label{fig5}
\end{figure}

Fig.~\ref{fig5} shows the optical images of BS1--BS4 under the thin disk source. Four observer inclination angles are selected, $\theta_o=0^\circ,30^\circ,60^\circ,80^\circ$, with the field of view fixed at $\gamma_{\mathrm{fov}}=3^\circ$. The boson star is located at the center of each image, and the colorbar reflects the variation in intensity, where white indicates the maximum intensity and purple indicates the minimum. When $\theta_o=0^\circ$ (the first column), a circular bright ring appears in the image, corresponding to the direct image. In this case, since the line of sight is perpendicular to the equatorial plane, gravitational redshift dominates. As $\phi_0$ increases, the shape of the direct image remains unchanged, but its overall size decreases. When $\theta_o$ increases to $30^\circ$ (the second column), Doppler redshift due to relative motion becomes more significant, resulting in an asymmetric direct image with greater intensity on the left side than the right. Under these conditions, for larger initial scalar field ($\phi_0=0.6$), a bright crescent-shaped pattern appears in the image, corresponding to the lensed image. As $\theta_o$ further increases (the third and fourth columns), the lensed image becomes more distinguishable and deforms into a D-shaped pattern, while the direct image gradually evolves into a cap-like shape. These results indicate that the observer inclination angle $\theta_o$ affects the overall shape of the optical image, while the initial scalar field $\phi_0$ influences the size of the image by changing the compactness of the star. For larger $\phi_0$, it is more likely that a lensed image appears in the observation.

\begin{figure}[!h]
	\centering 
	
	\subfigure[$\mathcal{G}=0.01,\theta_o=0^\circ$]{\includegraphics[scale=0.35]{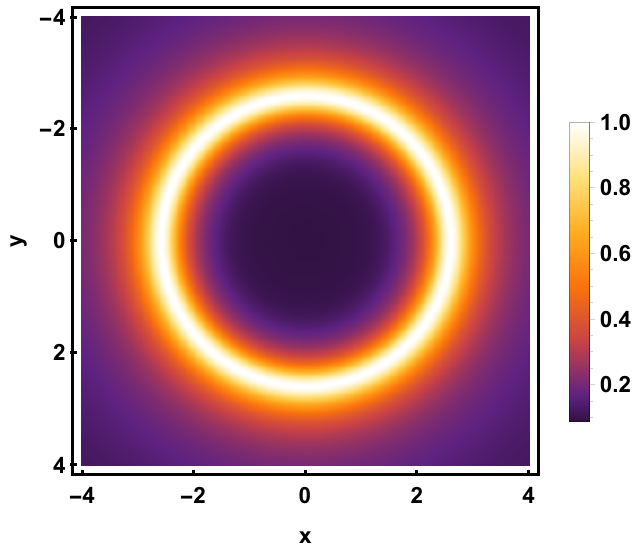}}
	\subfigure[$\mathcal{G}=0.01,\theta_o=30^\circ$]{\includegraphics[scale=0.35]{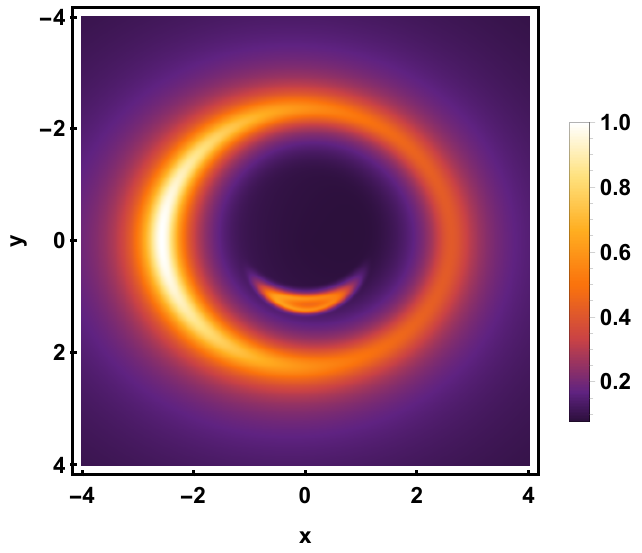}}
	\subfigure[$\mathcal{G}=0.01,\theta_o=60^\circ$]{\includegraphics[scale=0.35]{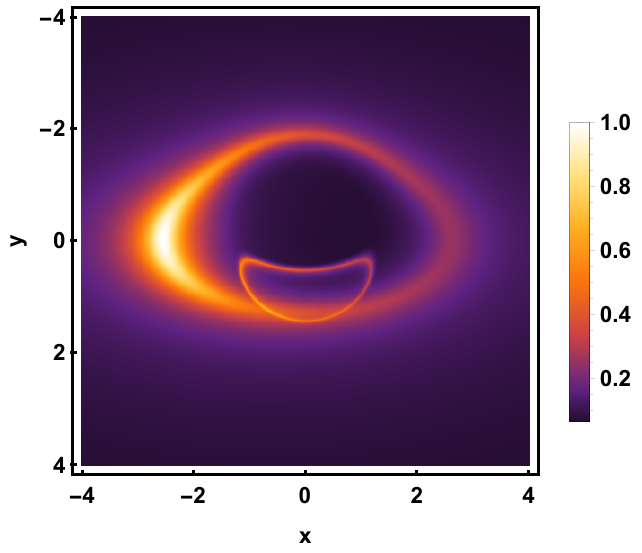}}
	\subfigure[$\mathcal{G}=0.01,\theta_o=80^\circ$]{\includegraphics[scale=0.35]{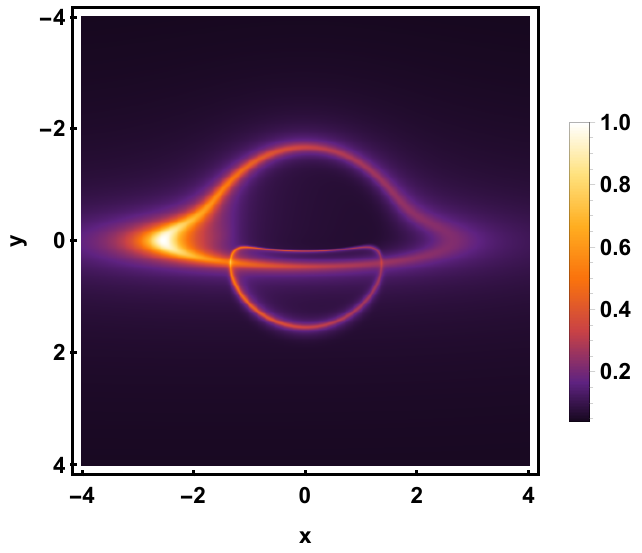}}
	
	\subfigure[$\mathcal{G}=0.07,\theta_o=0^\circ$]{\includegraphics[scale=0.35]{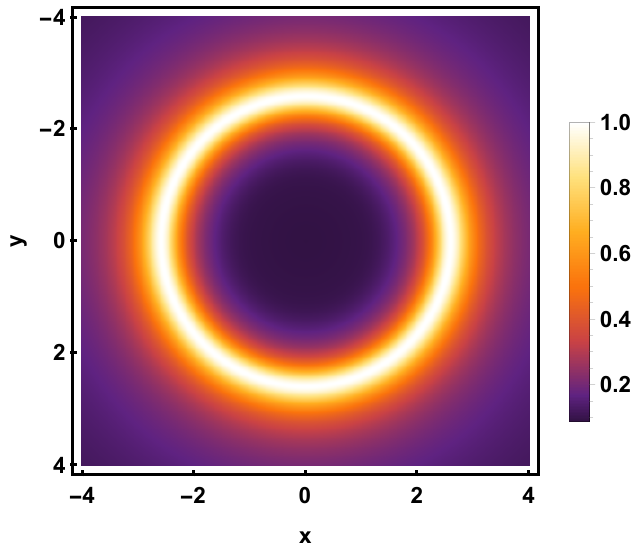}}
	\subfigure[$\mathcal{G}=0.07,\theta_o=30^\circ$]{\includegraphics[scale=0.35]{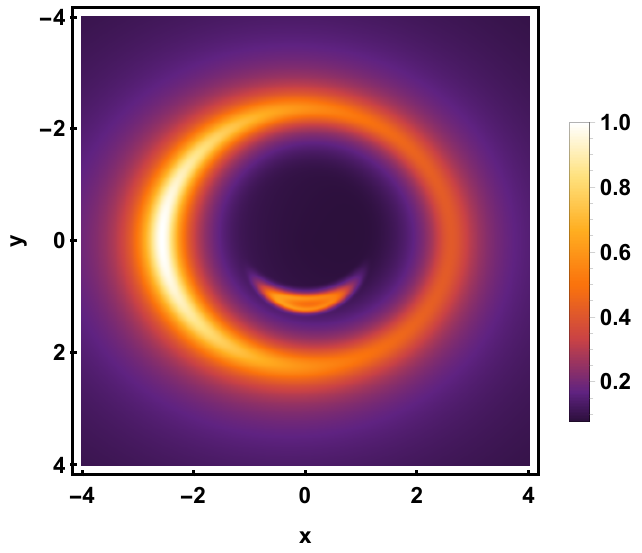}}
	\subfigure[$\mathcal{G}=0.07,\theta_o=60^\circ$]{\includegraphics[scale=0.35]{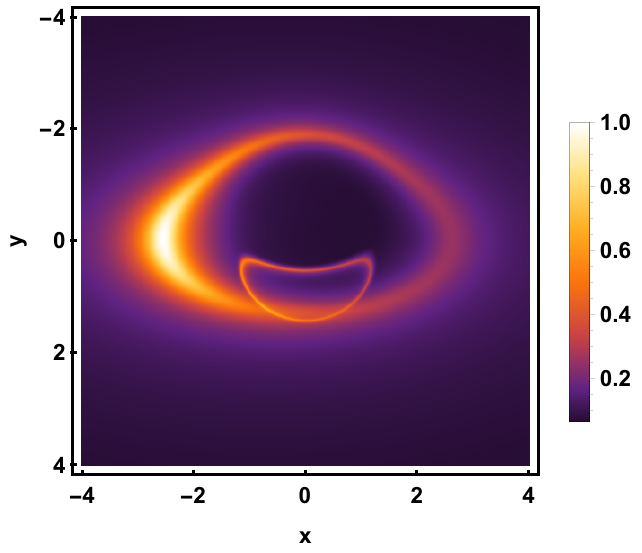}}
	\subfigure[$\mathcal{G}=0.07,\theta_o=80^\circ$]{\includegraphics[scale=0.35]{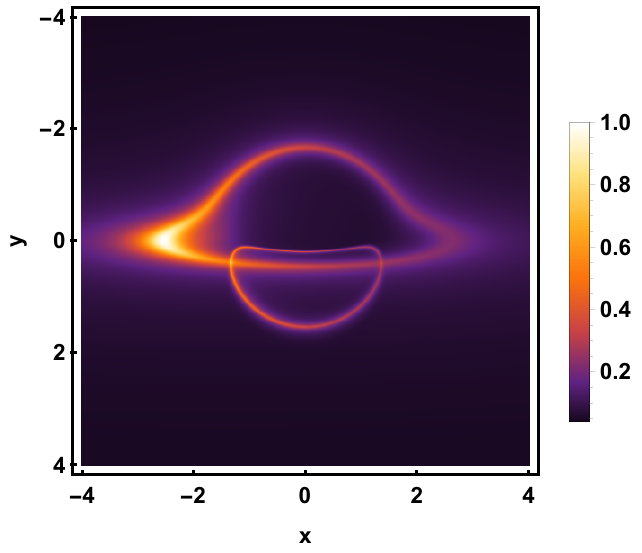}}
	
	\subfigure[$\mathcal{G}=0.1,\theta_o=0^\circ$]{\includegraphics[scale=0.35]{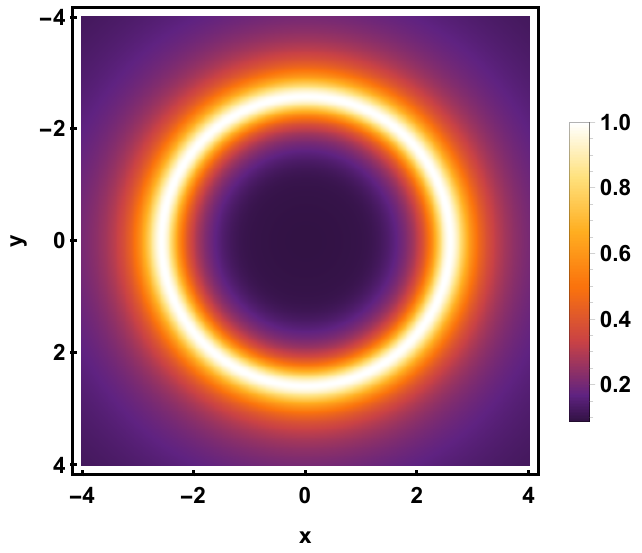}}
	\subfigure[$\mathcal{G}=0.1,\theta_o=30^\circ$]{\includegraphics[scale=0.35]{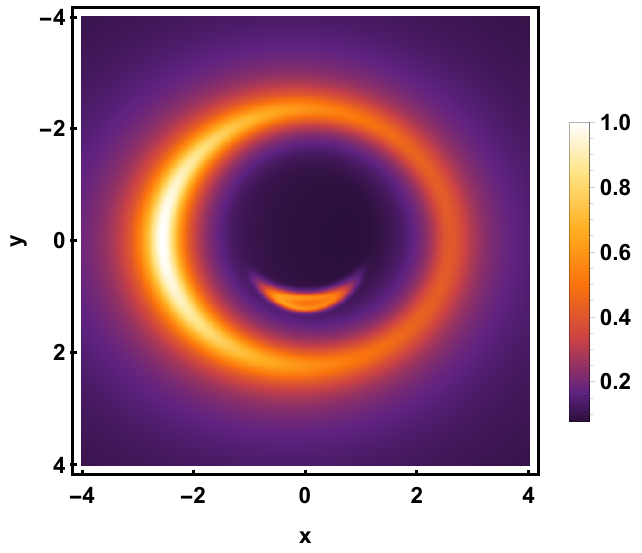}}
	\subfigure[$\mathcal{G}=0.1,\theta_o=60^\circ$]{\includegraphics[scale=0.35]{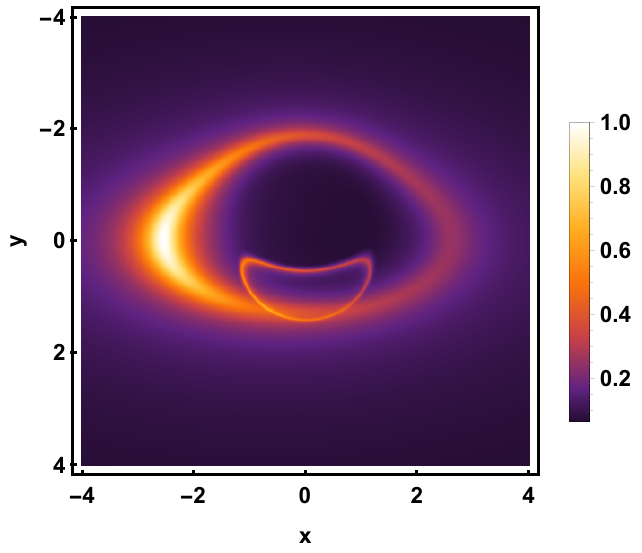}}
	\subfigure[$\mathcal{G}=0.1,\theta_o=80^\circ$]{\includegraphics[scale=0.35]{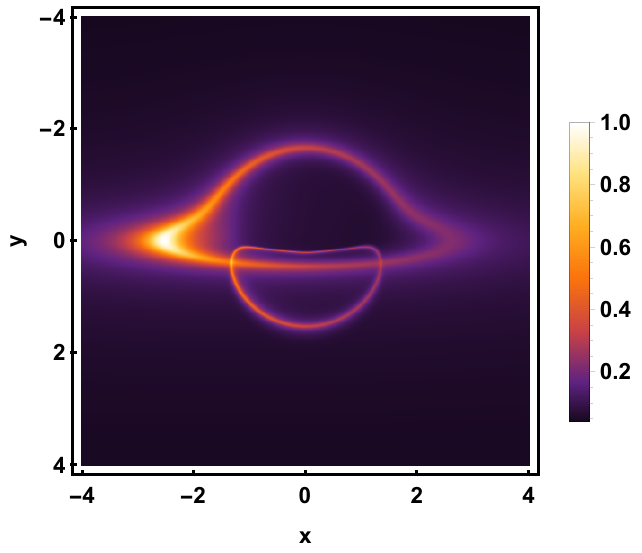}}
	
	\subfigure[$\mathcal{G}=0.15,\theta_o=0^\circ$]{\includegraphics[scale=0.35]{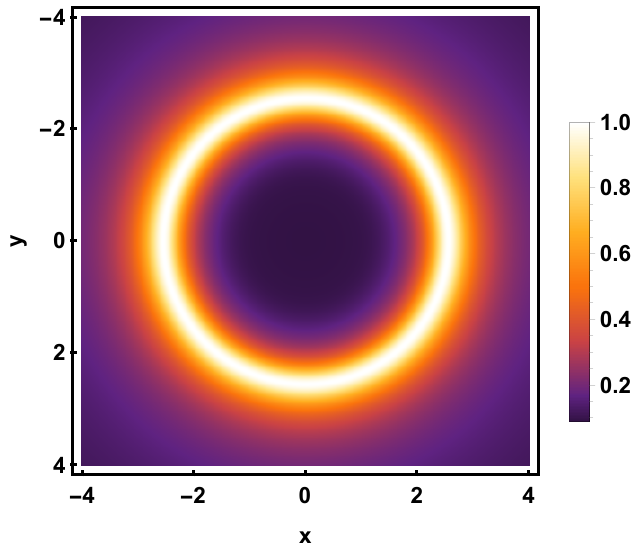}}
	\subfigure[$\mathcal{G}=0.15,\theta_o=30^\circ$]{\includegraphics[scale=0.35]{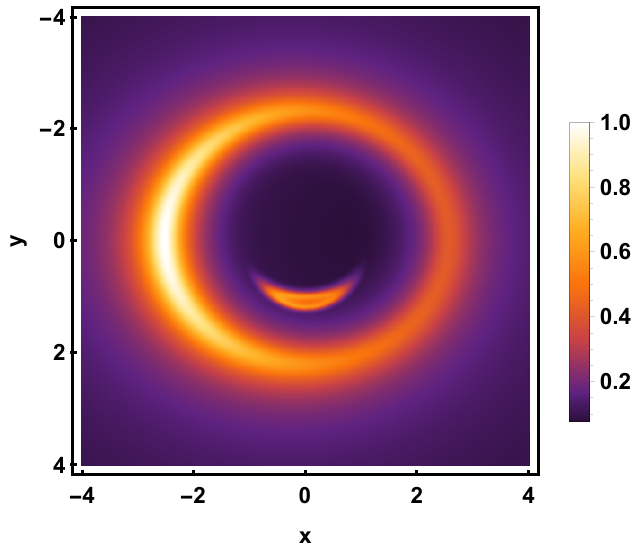}}
	\subfigure[$\mathcal{G}=0.15,\theta_o=60^\circ$]{\includegraphics[scale=0.35]{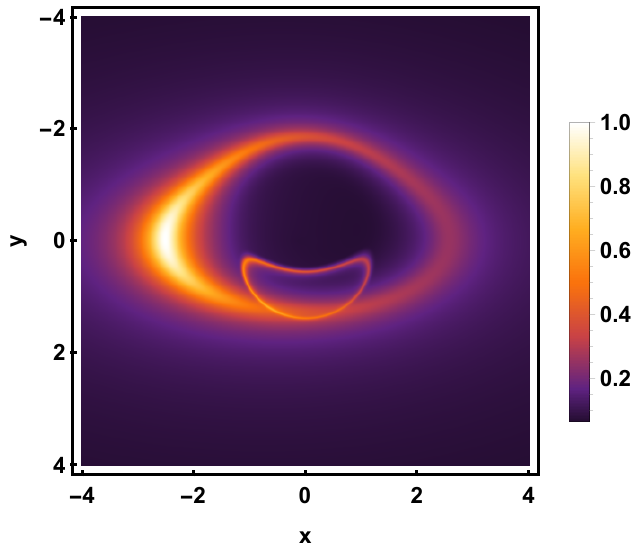}}
	\subfigure[$\mathcal{G}=0.15,\theta_o=80^\circ$]{\includegraphics[scale=0.35]{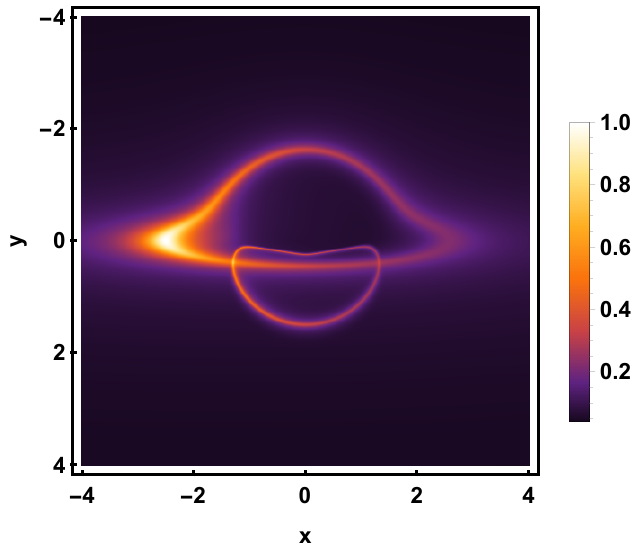}}
	
	\caption{Optical images of boson stars under the thin accretion disk. Each row from top to bottom corresponds to magnetic charges $\mathcal{G}=0.01,0.07,0.1,0.15$, and each column from left to right corresponds to observer inclination angles $\theta_o=0^\circ, 30^\circ, 60^\circ, 80^\circ$. The fixed parameters are $\mathcal{S}=0.2,\phi_0=0.6,\gamma_{\mathrm{fov}}=2.7^\circ$.}
	\label{fig6}
\end{figure}

Fig.~\ref{fig6} shows the optical images of BS5--BS8 under the thin disk source, with the field of view fixed at $\gamma_{\mathrm{fov}}=2.7^\circ$. When $\theta_o=0^\circ$ (the first column), only the direct image appears. When $\theta_o$ increases to $30^\circ$ (the second column), lensed images appear in all cases. As $\theta_o$ further increases, the lensed images become more clearly visible. The parameter $\mathcal{G}$ mainly affects the size of the optical images; as $\mathcal{G}$ increases, the sizes of both the direct and lensed images decrease slightly. It is noteworthy that, although boson stars do not possess an event horizon, for both Fig.~\ref{fig5} and Fig.~\ref{fig6}, a region of reduced intensity always appears at the center of the image regardless of the parameter values, similar to the inner shadow of a black hole. This suggests that it may be difficult to distinguish boson stars from black holes based solely on optical images.

\begin{figure}[!h]
	\centering 
	
	\subfigure[$\theta_o=0^\circ$]{\includegraphics[scale=0.35]{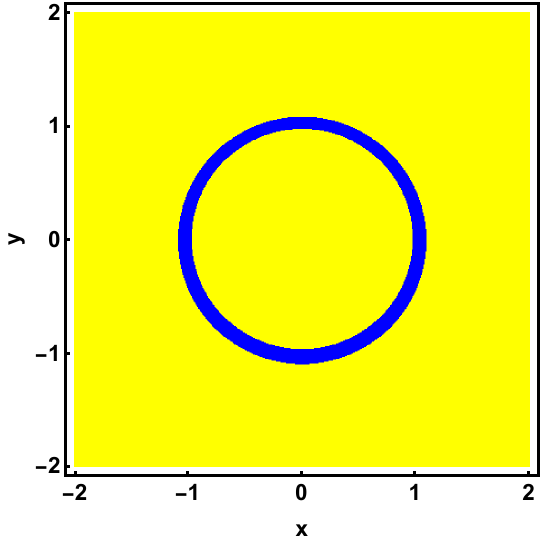}}
	\subfigure[$\theta_o=30^\circ$]{\includegraphics[scale=0.35]{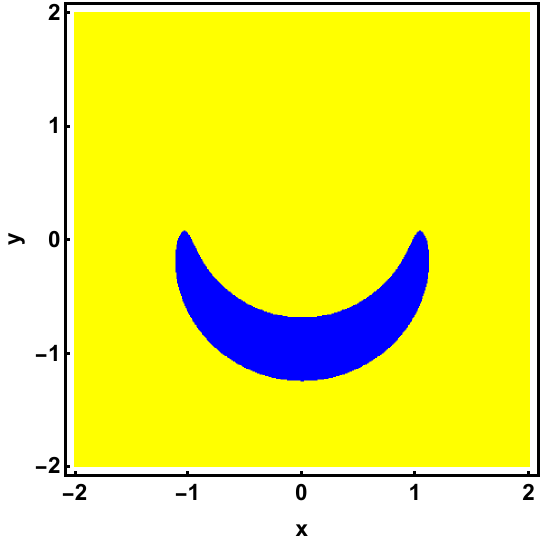}}
	\subfigure[$\theta_o=60^\circ$]{\includegraphics[scale=0.35]{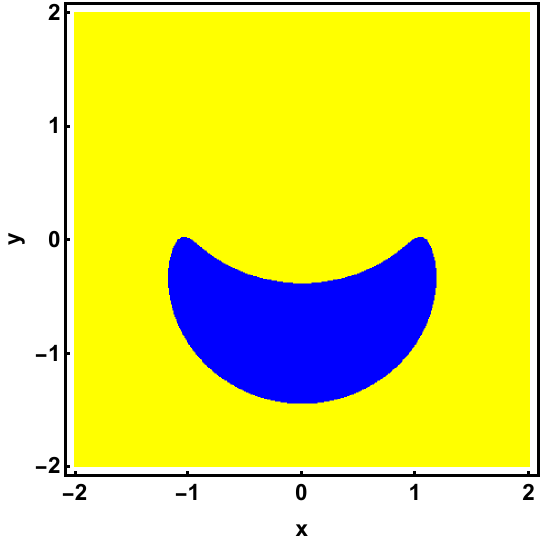}}
	\subfigure[$\theta_o=80^\circ$]{\includegraphics[scale=0.35]{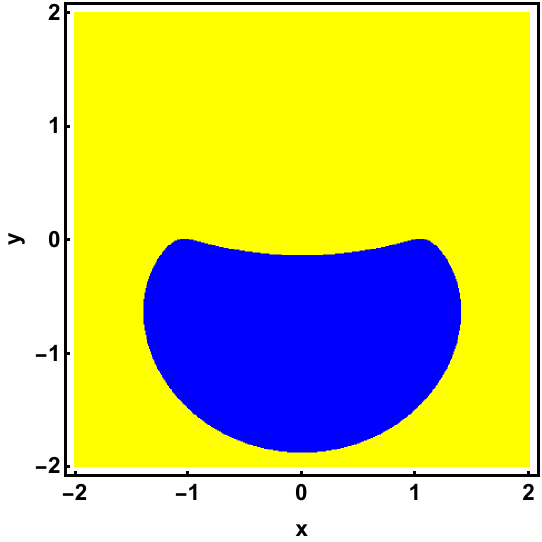}}
	
	\caption{The lensing bands for boson stars under a prograde thin accretion disk. The yellow and blue regions indicate photons that cross the equatorial plane once and twice, corresponding to the direct and lensed images, respectively. The fixed parameters are $\mathcal{S} = 0.8$, $\mathcal{G} = 0.35$, $\phi_0=0.6$, and $\gamma_{\mathrm{fov}} = 1.5^\circ$.}
	\label{fig7}
\end{figure}

\begin{figure}[!h]
	\centering 
	
	\subfigure[The red, green, blue, and orange curves correspond to BS1--BS4, respectively.]{\includegraphics[scale=0.5]{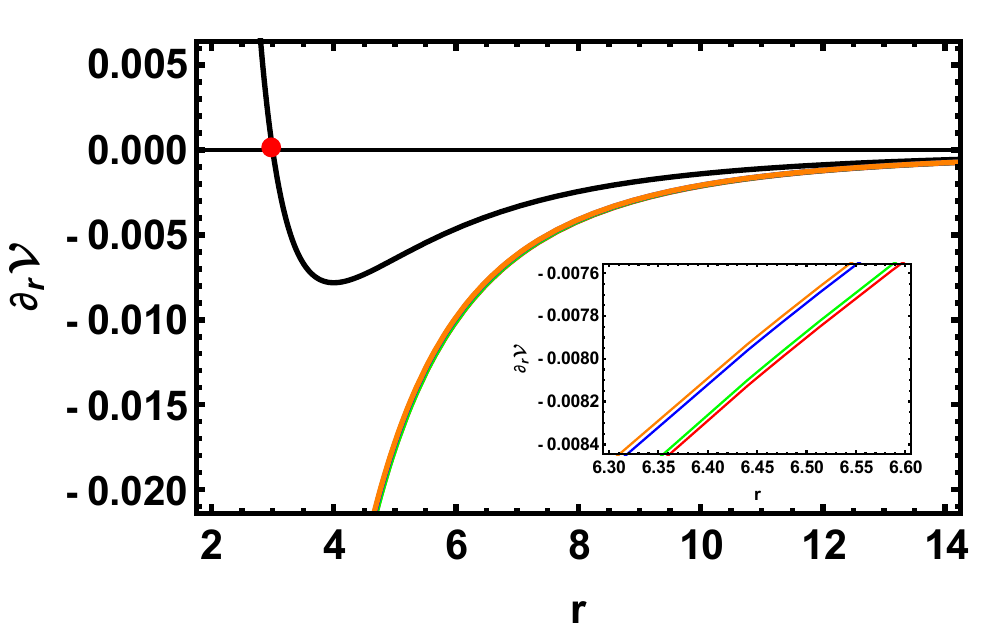}}
	\subfigure[The red, green, blue, and orange curves correspond to BS5--BS8, respectively.]{\includegraphics[scale=0.5]{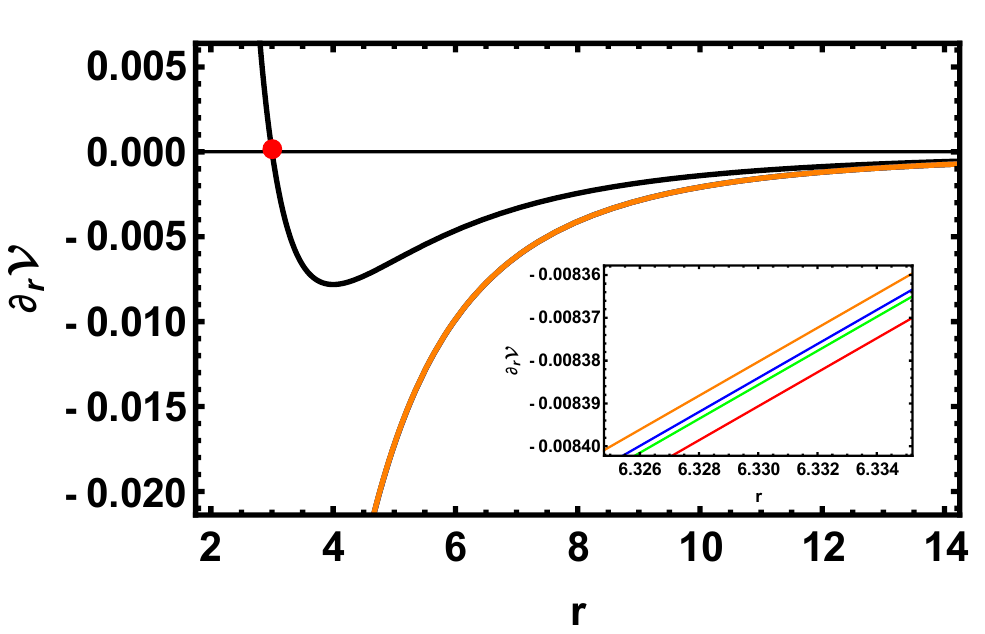}}
	
	\caption{First derivative of the effective potential $\mathcal{V}_{photon}(r)$. }
\label{fig8}
\end{figure}

To more intuitively distinguish between the direct and lensed images, Fig.~\ref{fig7} presents the lensing bands for different observer inclination angles. In the figure, yellow and blue represent photons that cross the equatorial plane once and twice, corresponding to the direct and lensed images, respectively. It can be seen that as $\theta_o$ increases, the lensed image gradually evolves from a ring shape to a crescent shape and further into a D-shaped pattern. Interestingly, for all images, only direct and lensed images appear on the imaging plane. This indicates that, for a given field of view, all light rays within the field of view in the boson star spacetime will intersect the accretion disk when traced backward. In contrast, for black holes, due to the presence of the event horizon, some light rays are directly absorbed and cannot intersect the accretion disk. On the other hand, no higher-order images appear in Fig.~\ref{fig7}, which seems to suggest that photon rings do not exist in the boson star spacetime. To further determine whether photon rings exist in the spacetimes of BS1--BS4 and BS5--BS8, Fig.~\ref{fig8} presents the first derivatives of $\mathcal{V}_{photon}(r)$ under the condition $\mathcal{V}_{photon}(r)=0$, where $\mathcal{V}_{photon}(r)$ denotes the effective potential defined in Eq.~(\ref{eq:veffp}).

In Fig.~\ref{fig8}, the black solid line represents the first derivative of the effective potential for the Schwarzschild black hole with unit mass. It can be seen that there is a zero at $r = 3$, indicating that the effective potential has an extremum at this point, which precisely corresponds to the location of the photon ring of the Schwarzschild black hole. However, for BS1--BS4 and BS5--BS8, the first derivatives of the effective potential do not have any zeros, implying that there are no photon rings in the boson star spacetimes discussed in this paper. On the other hand, according to Ref.~\cite{huang2025orbits}, when the magnetic charge $\mathcal{G}$ takes larger values (for example, $\mathcal{G} > 0.6$), the Bardeen–boson star enters a so-called “frozen” phase, referred to as a frozen Bardeen–boson star. A frozen Bardeen–boson star exhibits black hole–like properties, such as the presence of a critical horizon that closely resembles an event horizon. Because the spacetime of a frozen Bardeen–boson star is extremely compact, the effective potential develops an extremum, leading to the formation of a photon ring. The emergence of a photon ring can therefore be regarded as a geometric indicator of the transition from a Bardeen–boson star to its frozen state.


\section{Polarized images of boson stars}\label{sec4}
In this section, we study the polarization images of boson stars based on the thin accretion disk model, aiming to provide a more comprehensive understanding of the radiation characteristics of boson stars and their interactions with the surrounding environment. It is assumed that the polarized radiation originates from synchrotron emission by electrons in the plasma. For an observer comoving with the plasma, whose four-velocity is $u^\mu$, the polarization direction $\vec{f}$ of the emitted light is always perpendicular to both the local magnetic field $\vec{\mathcal{B}}$ and the wave vector $\vec{\mathcal{K}}$ of the photon,
\begin{equation}
	\vec{f}=\frac{\vec{\mathcal{K}}\times\vec{\mathcal{B}}}{|\vec{\mathcal{K}}||\vec{\mathcal{B}}|},
\end{equation}
and the general covariant form of the above equation can be written as
\begin{equation}
	f^\mu \propto \epsilon^{\mu\nu\alpha\beta}u_\nu\mathcal{K}_\alpha\mathcal{B}_\beta.
\end{equation}
During the imaging process, it is necessary to first determine the direction of the polarization vector and normalize it so that it satisfies the orthonormal condition
\begin{equation}
	f^{\mu}f_{\mu}=1.
\end{equation}
The intensities of linearly polarized and natural light at the emission point are represented by the emission functions $e_p$ and $e_i$, respectively. To simplify the calculation, it is assumed that the emission intensity is independent of photon frequency and magnetic field, and depends only on the position, so that
\begin{equation}
	e_i=e_i(r),\qquad e_p=\xi e_i(r),
\end{equation}
where $\xi\in[0,1]$ is a function describing the proportion of linearly polarized light in the total intensity at the emission point. If it is assumed that the emitted light is completely linearly polarized, then $\xi=1$. In the geometric optics approximation, the polarization vector $f^\mu$ undergoes parallel transport along the geodesic of the photon,
\begin{equation}
	\mathcal{K}^{\nu}\nabla_{\nu}f^{\mu}=0.
\end{equation}
This equation can be rewritten as
\begin{equation}
	\frac{d}{d\lambda}f^{\mu}+\Gamma^{\mu}{}_{\nu\alpha}\mathcal{K}^{\nu}f^{\alpha}=0,
\end{equation}
where $\lambda$ is the affine parameter. At the observer, the intensity of linearly polarized light $\mathcal{P}_{\nu_o}$ and the total intensity $s_{\nu_o}$ are the same as in the unpolarized case,
\begin{equation}
	\mathcal{P}_{\nu_o}=\chi^3e_p,\qquad s_{\nu_o}=\chi^3e_i,
\end{equation}
where $\chi$ is the redshift factor. In (\ref{eq:tetrad}), a set of orthogonal tetrads (ZAMO) is established at the observer, and the imaging screen is set up accordingly. On the screen, a set of orthonormal vectors $(e_{(\theta)},\,e_{(\varphi)})$ is chosen as the basis, and the projection of the polarization vector onto the imaging plane can be expressed as
\begin{equation}
	f^{(\alpha)}=f^\mu\cdot e_\alpha=-f^\mu\cdot e_\varphi,\qquad f^{(\beta)}=f^\mu\cdot e_\beta=-f^\mu\cdot e_\theta.
\end{equation}
After determining the direction of the polarization vector, the total intensity of linearly polarized light can be obtained by summing the linearly polarized intensity at each emission point on the equatorial plane. According to the definitions of the Stokes parameters $\mathcal{Q}$ and $\mathcal{U}$~\cite{huang2024coport}, these parameters satisfy the principle of linear superposition, so the final results can be obtained by summing $\mathcal{Q}$ and $\mathcal{U}$,
\begin{equation}
	\mathcal{Q}_{all}=\sum_{n=1}^{N}\chi_n^3e_{pn}\left[\left(f_n^{(\alpha)}\right)^2-\left(f_n^{(\beta)}\right)^2\right],\quad
	\mathcal{U}_{all}=\sum_{n=1}^{N}\chi_n^3e_{pn}\left(2f_n^{(\alpha)}f_n^{(\beta)}\right),
\end{equation}
where $n=1,\ldots,N$ denotes the number of times the light ray crosses the equatorial plane. The total intensity of linearly polarized light and the electric vector position angle (EVPA) are given by
\begin{equation}
	\mathcal{P}_o=\sqrt{\mathcal{Q}_{all}^2+\mathcal{U}_{all}^2},\quad\Phi=\frac{1}{2}\arctan\frac{\mathcal{U}_{all}}{\mathcal{Q}_{all}}.
\end{equation}
We adopt the gauge $f^{(\beta)}>0,\Phi\in(0,\pi)$. Based on this and the thin accretion disk model in Sec.~\ref{sec3}, the intensity and direction of polarized light at the observer can be calculated.

\begin{figure}[!h]
	\centering 
	\subfigure[$\phi_0=0.3,\theta_o=0^\circ$]{\includegraphics[scale=0.35]{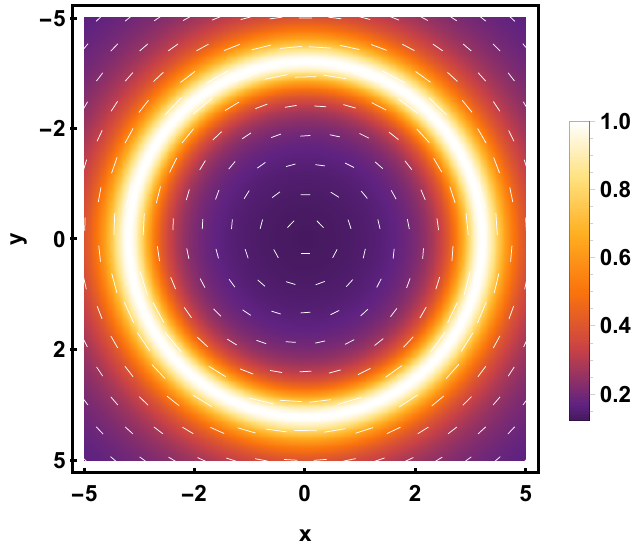}}
	\subfigure[$\phi_0=0.3,\theta_o=30^\circ$]{\includegraphics[scale=0.35]{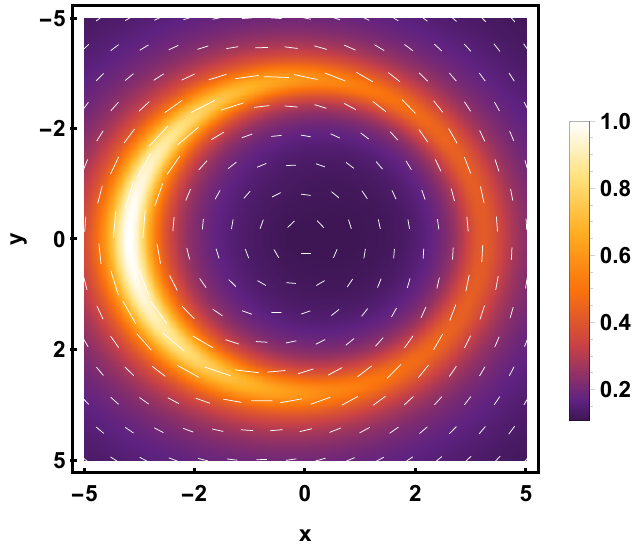}\label{fig9b}}
	\subfigure[$\phi_0=0.3,\theta_o=60^\circ$]{\includegraphics[scale=0.35]{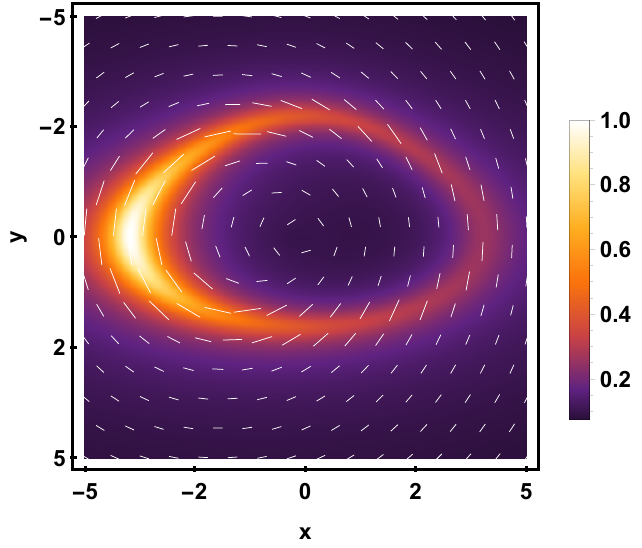}}
	\subfigure[$\phi_0=0.3,\theta_o=80^\circ$]{\includegraphics[scale=0.35]{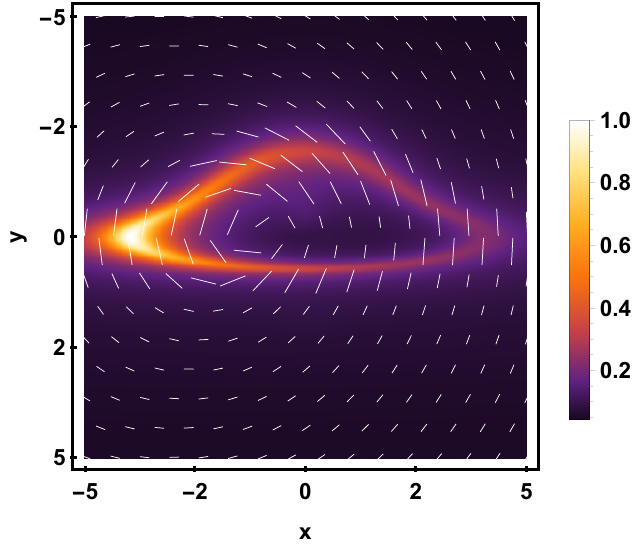}}
	
	\subfigure[$\phi_0=0.35,\theta_o=0^\circ$]{\includegraphics[scale=0.35]{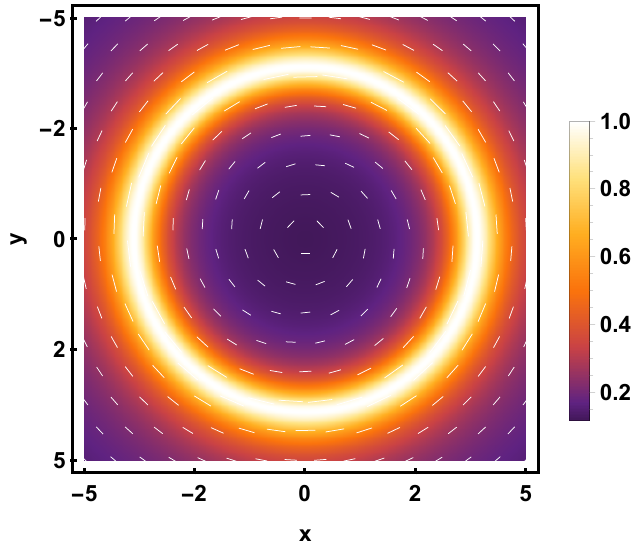}}
	\subfigure[$\phi_0=0.35,\theta_o=30^\circ$]{\includegraphics[scale=0.35]{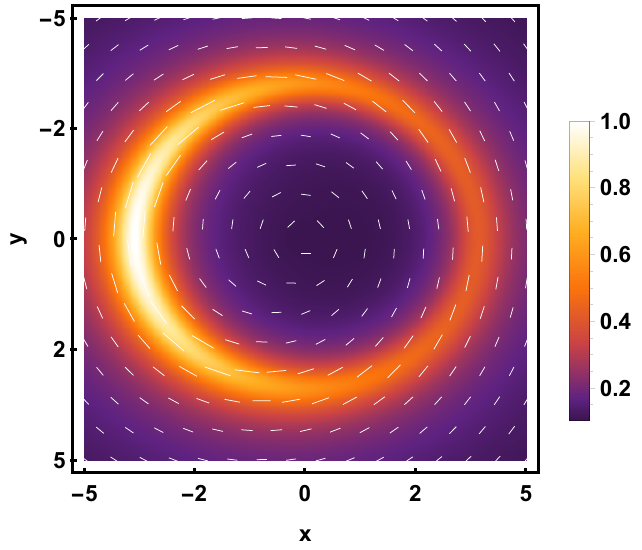}}
	\subfigure[$\phi_0=0.35,\theta_o=60^\circ$]{\includegraphics[scale=0.35]{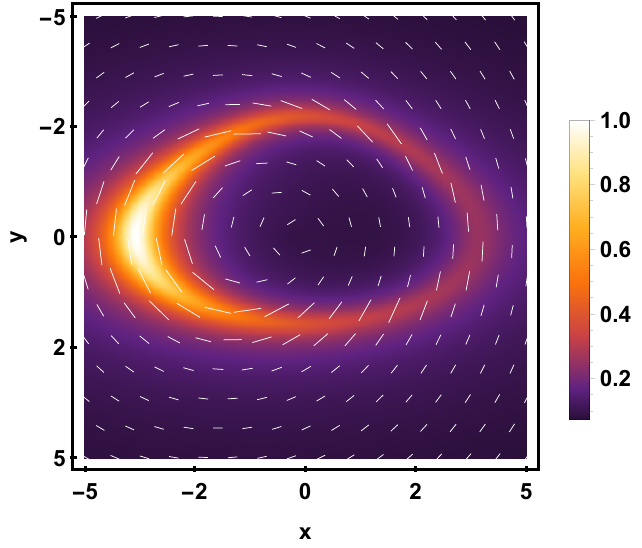}}
	\subfigure[$\phi_0=0.35,\theta_o=80^\circ$]{\includegraphics[scale=0.35]{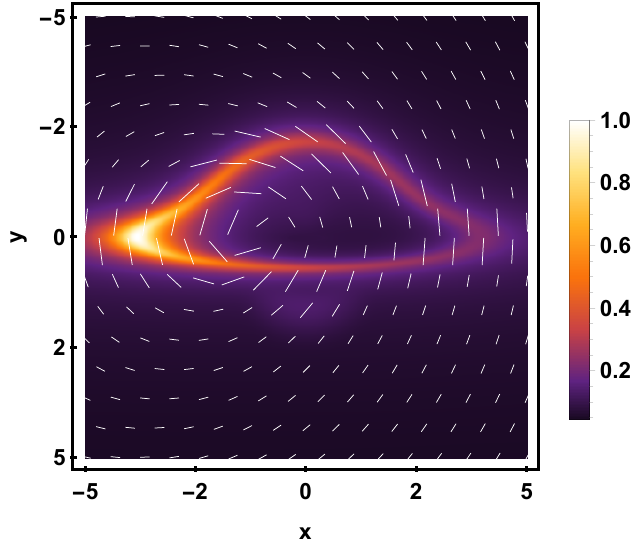}}
	
	\subfigure[$\phi_0=0.55,\theta_o=0^\circ$]{\includegraphics[scale=0.35]{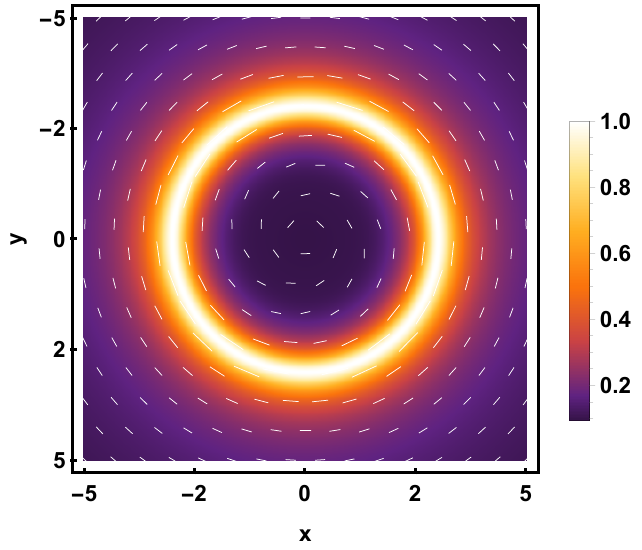}}
	\subfigure[$\phi_0=0.55,\theta_o=30^\circ$]{\includegraphics[scale=0.35]{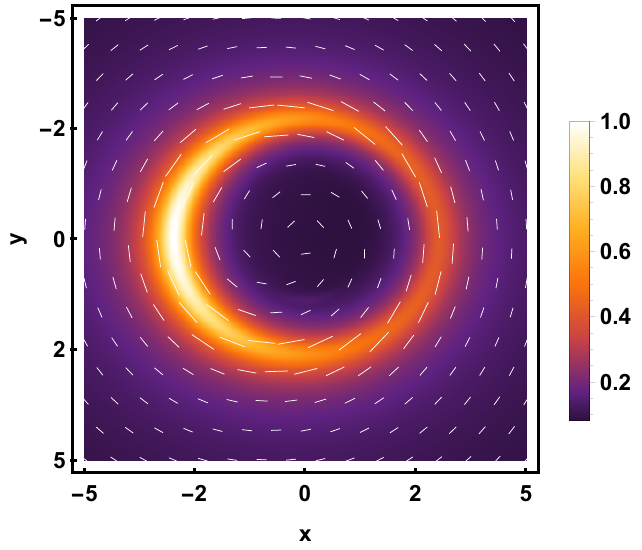}}
	\subfigure[$\phi_0=0.55,\theta_o=60^\circ$]{\includegraphics[scale=0.35]{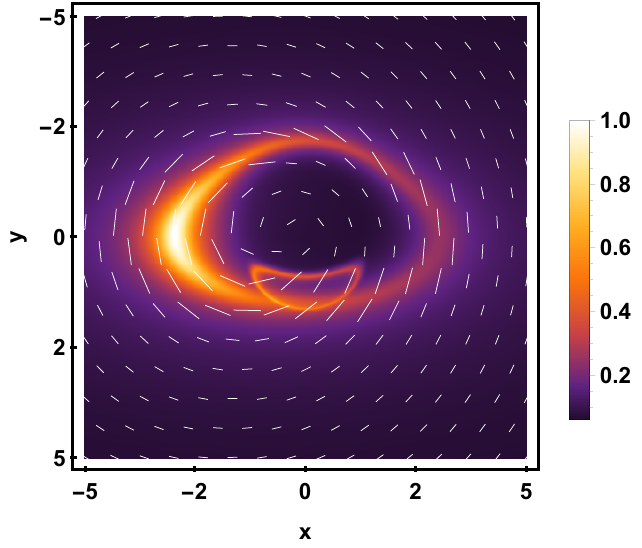}}
	\subfigure[$\phi_0=0.55,\theta_o=80^\circ$]{\includegraphics[scale=0.35]{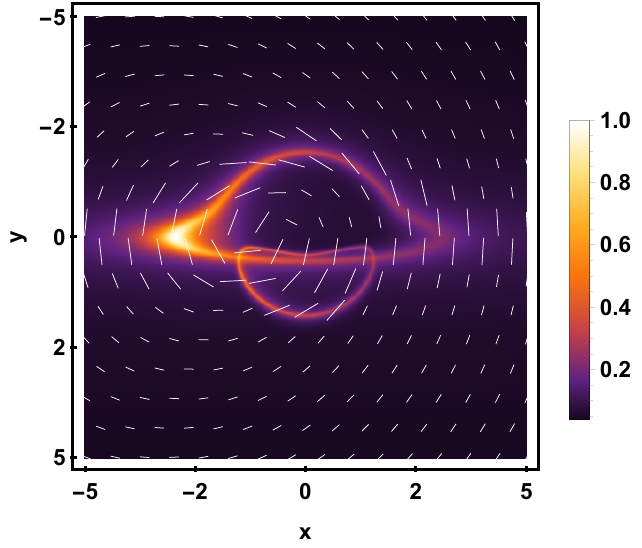}}
	
	\subfigure[$\phi_0=0.6,\theta_o=0^\circ$]{\includegraphics[scale=0.35]{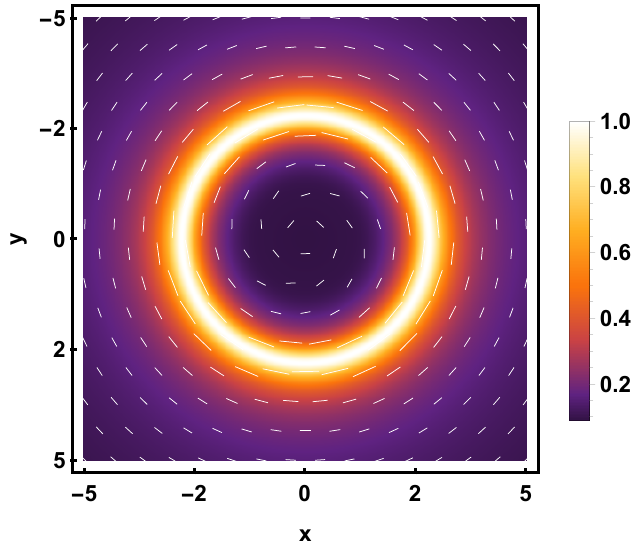}}
	\subfigure[$\phi_0=0.6,\theta_o=30^\circ$]{\includegraphics[scale=0.35]{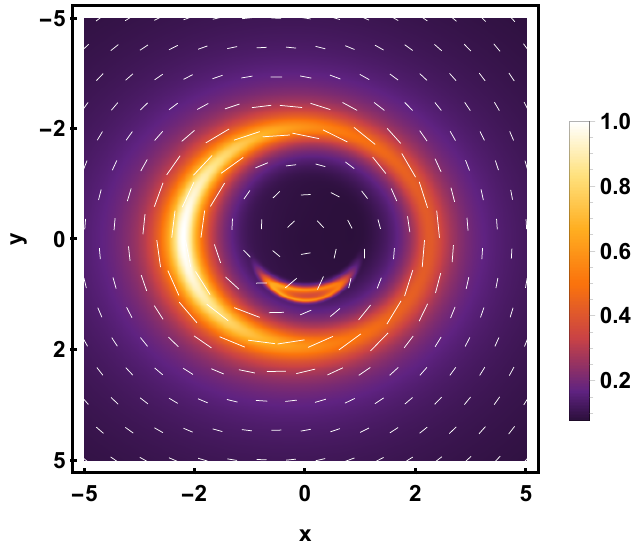}}
	\subfigure[$\phi_0=0.6,\theta_o=60^\circ$]{\includegraphics[scale=0.35]{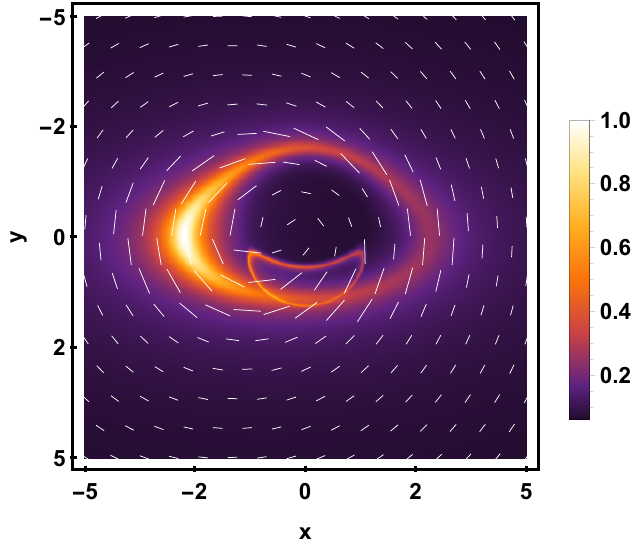}}
	\subfigure[$\phi_0=0.6,\theta_o=80^\circ$]{\includegraphics[scale=0.35]{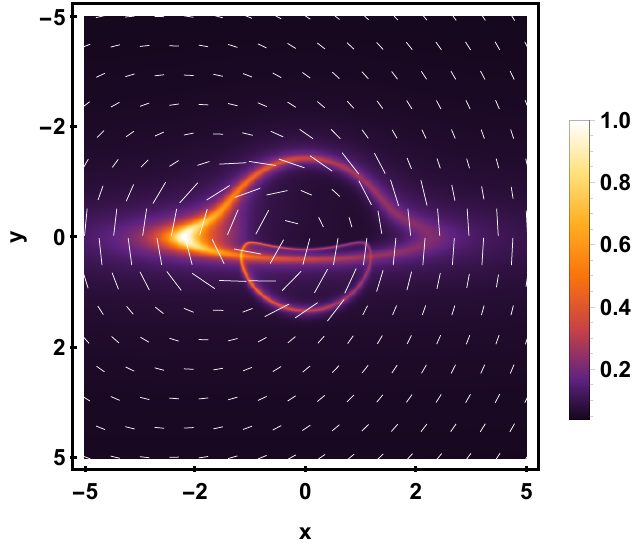}}
	
	\caption{Polarization images of the Bardeen-boson star generated by synchrotron radiation. White line segments indicate linear polarization vectors with varying intensities. All other parameters are fixed, with magnetic field $\vec{\mathcal{B}} = (0.87, 0.5, 0)$, $\mathcal{S} = 0.8$, $\mathcal{G} = 0.35$, and $\gamma_{\mathrm{fov}} = 3^\circ$.}
	\label{fig9}
\end{figure}

\begin{figure}[!h]
	\centering 
	
	\subfigure[$\mathcal{G}=0.01,\theta_o=0^\circ$]{\includegraphics[scale=0.35]{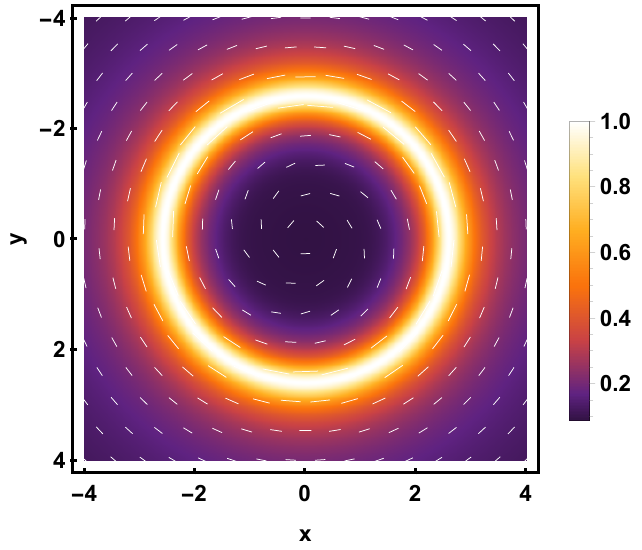}}
	\subfigure[$\mathcal{G}=0.01,\theta_o=30^\circ$]{\includegraphics[scale=0.35]{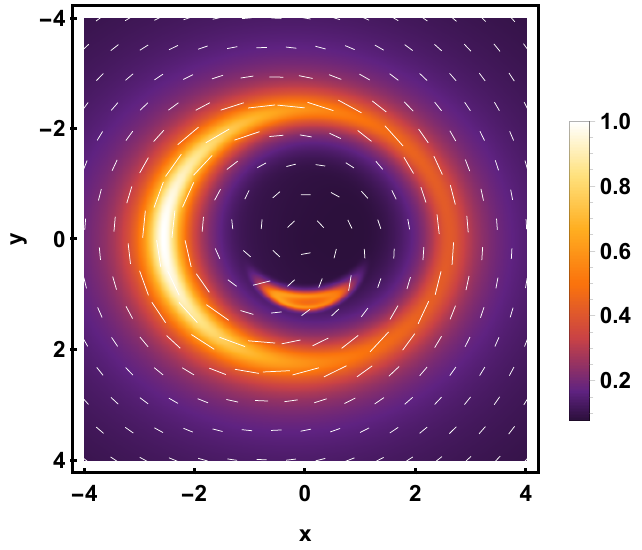}}
	\subfigure[$\mathcal{G}=0.01,\theta_o=60^\circ$]{\includegraphics[scale=0.35]{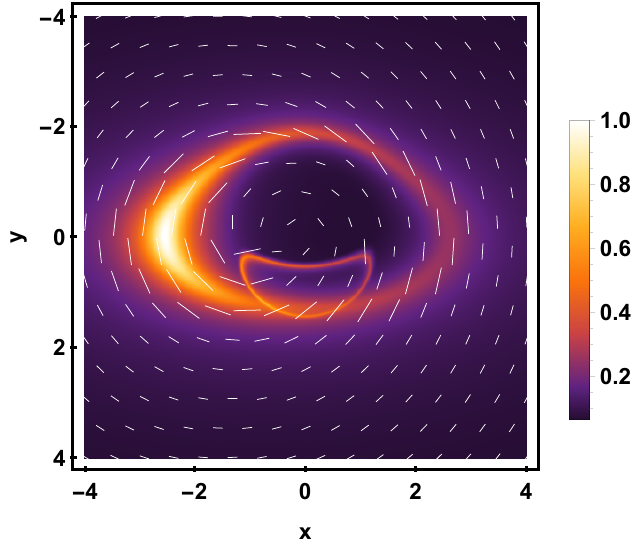}}
	\subfigure[$\mathcal{G}=0.01,\theta_o=80^\circ$]{\includegraphics[scale=0.35]{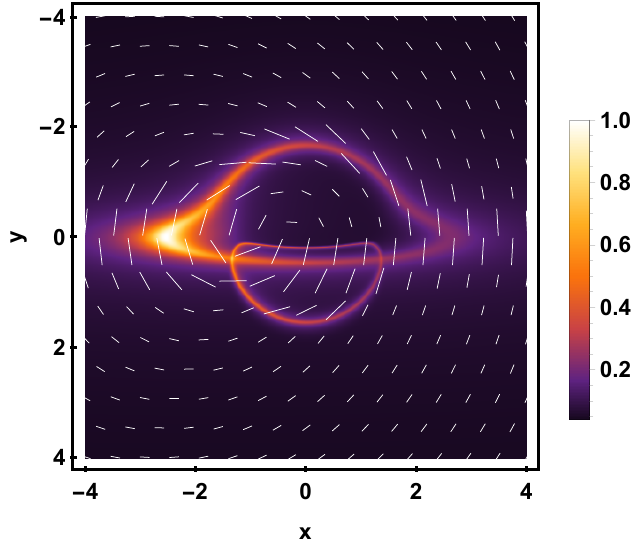}}
	
	\subfigure[$\mathcal{G}=0.07,\theta_o=0^\circ$]{\includegraphics[scale=0.35]{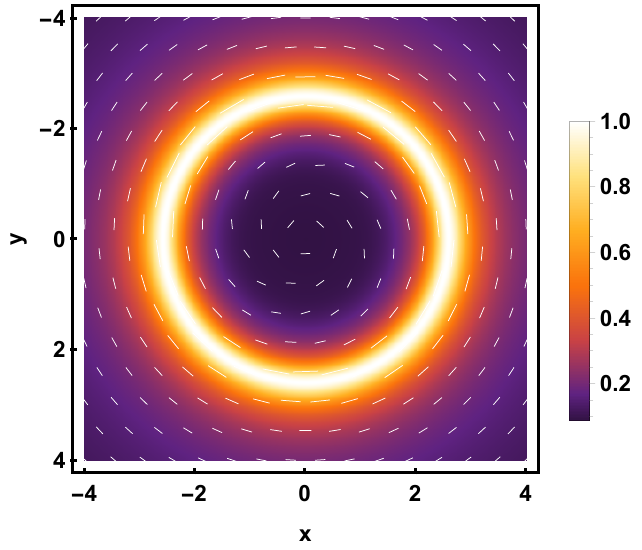}}
	\subfigure[$\mathcal{G}=0.07,\theta_o=30^\circ$]{\includegraphics[scale=0.35]{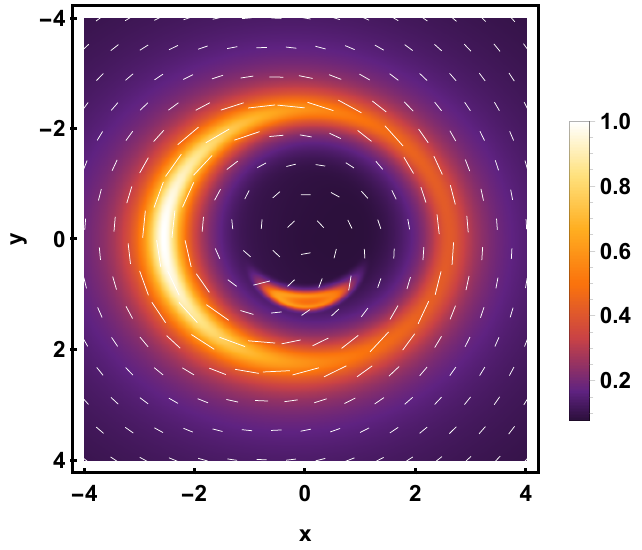}}
	\subfigure[$\mathcal{G}=0.07,\theta_o=60^\circ$]{\includegraphics[scale=0.35]{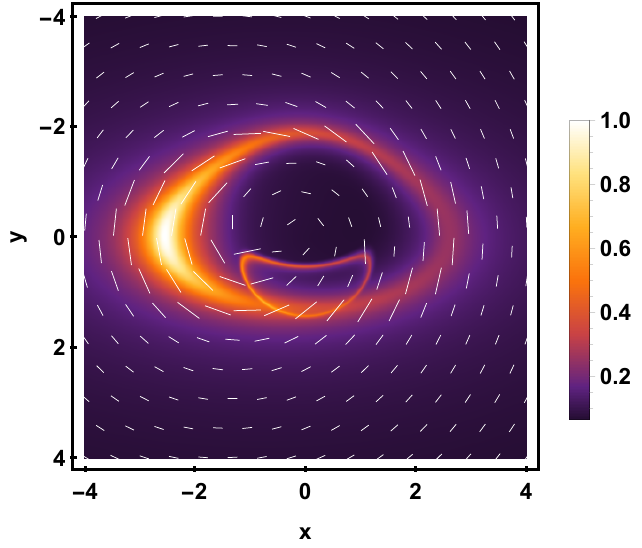}}
	\subfigure[$\mathcal{G}=0.07,\theta_o=80^\circ$]{\includegraphics[scale=0.35]{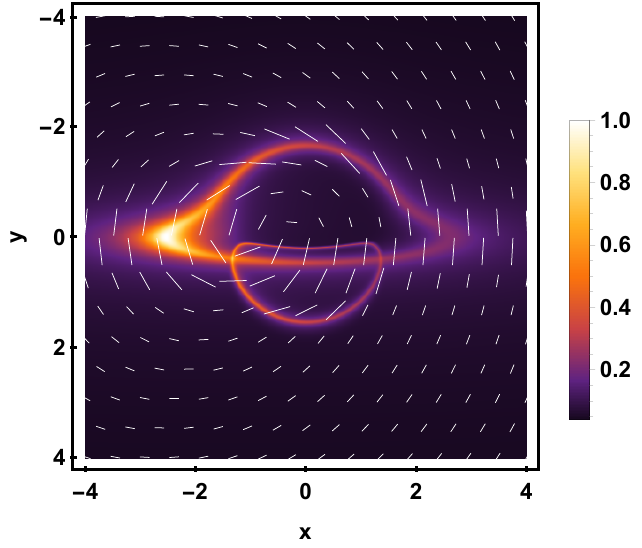}}
	
	\subfigure[$\mathcal{G}=0.1,\theta_o=0^\circ$]{\includegraphics[scale=0.35]{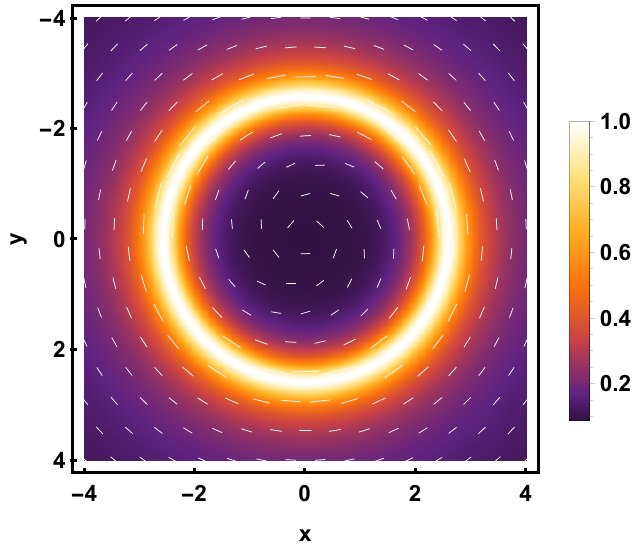}}
	\subfigure[$\mathcal{G}=0.1,\theta_o=30^\circ$]{\includegraphics[scale=0.35]{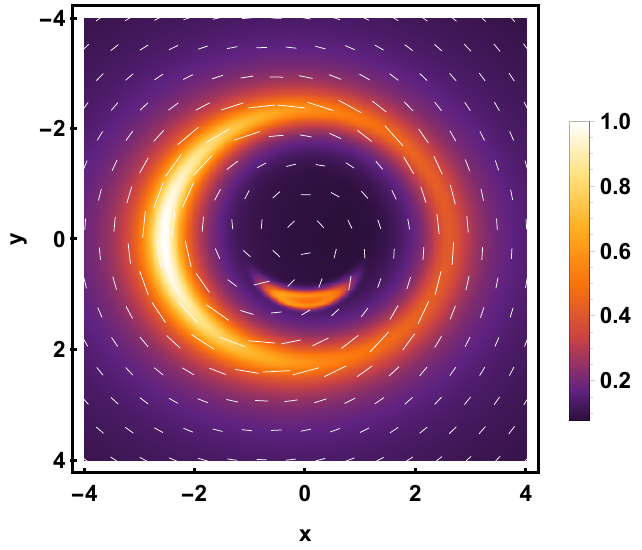}}
	\subfigure[$\mathcal{G}=0.1,\theta_o=60^\circ$]{\includegraphics[scale=0.35]{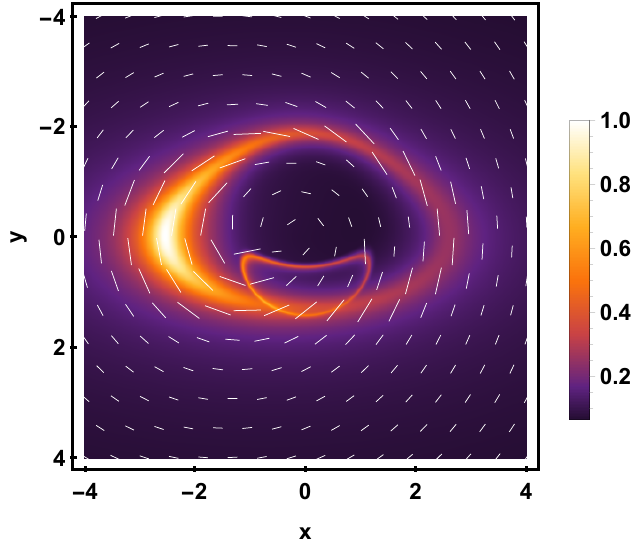}}
	\subfigure[$\mathcal{G}=0.1,\theta_o=80^\circ$]{\includegraphics[scale=0.35]{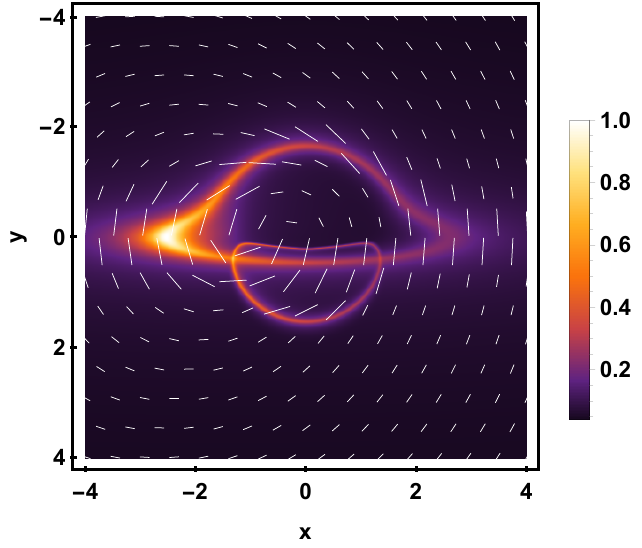}}
	
	\subfigure[$\mathcal{G}=0.15,\theta_o=0^\circ$]{\includegraphics[scale=0.35]{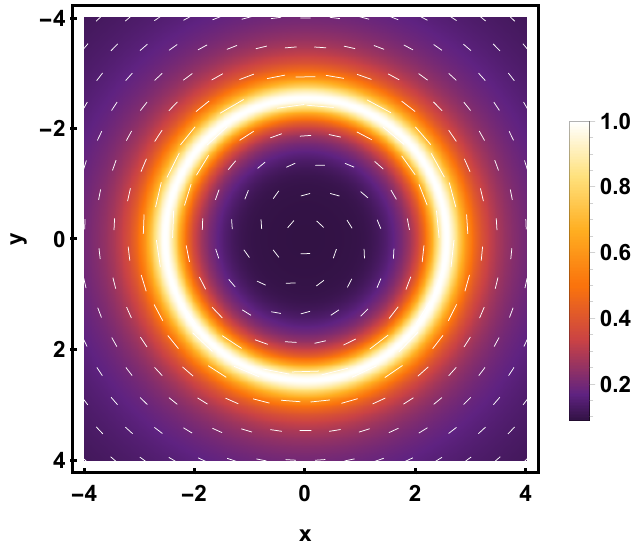}}
	\subfigure[$\mathcal{G}=0.15,\theta_o=30^\circ$]{\includegraphics[scale=0.35]{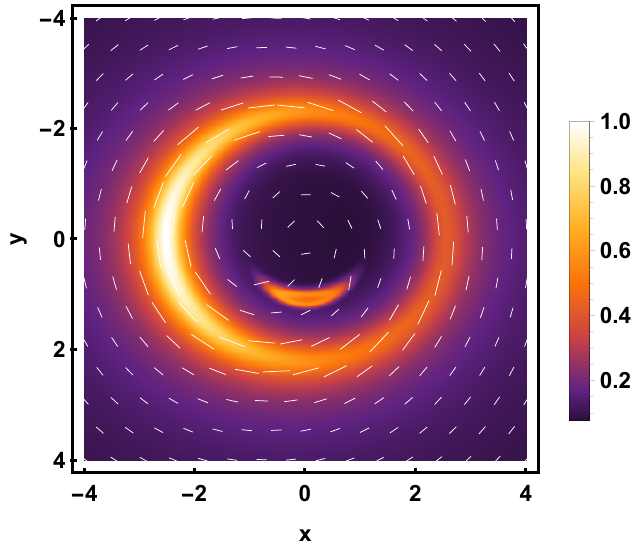}}
	\subfigure[$\mathcal{G}=0.15,\theta_o=60^\circ$]{\includegraphics[scale=0.35]{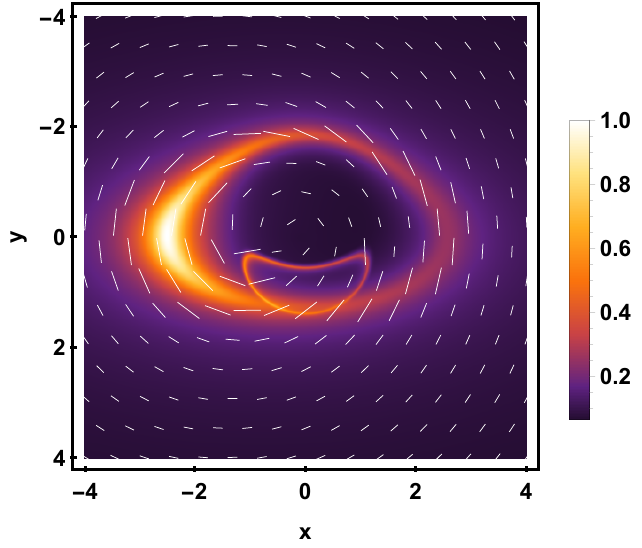}}
	\subfigure[$\mathcal{G}=0.15,\theta_o=80^\circ$]{\includegraphics[scale=0.35]{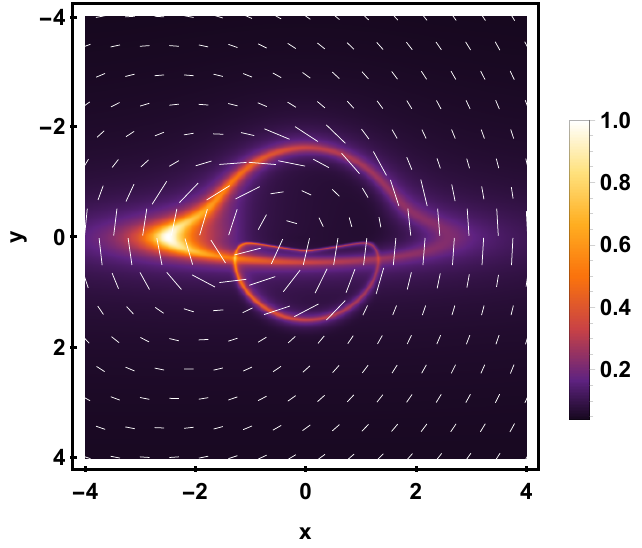}}
	
	\caption{Polarization images of the Bardeen-boson star generated by synchrotron radiation. White line segments indicate linear polarization vectors with varying intensities. All other parameters are fixed, with magnetic field $\vec{\mathcal{B}} = (0.87, 0.5, 0)$, $\mathcal{S}=0.2$, $\phi_0=0.6$, and $\gamma_{\mathrm{fov}}=2.7^\circ$.}
	\label{fig10}
\end{figure}

For M87*, assuming it is a Schwarzschild black hole, the optimal magnetic field configuration $\vec{\mathcal{B}}=(0.87,0.5,0)$ has been shown to effectively reproduce the observed polarization characteristics. Therefore, the same $\vec{\mathcal{B}}$ is adopted in the simulations of this paper. As in previous sections, four observer inclination angles $\theta_o=0^\circ,30^\circ,60^\circ,80^\circ$ are considered. Fig.~\ref{fig9} and~\ref{fig10} show the distributions of polarization vectors of BS1--BS4 and BS5--BS8, respectively. In the figures, white line segments represent the polarization vectors, whose lengths and directions reflect the total linearly polarized intensity $\mathcal{P}_o$ and the EVPA $\Phi$, respectively. The background of the images corresponds to thin accretion disk imaging, with specific details introduced earlier. It can be seen from the figures that the polarization intensity in the brighter regions is significantly enhanced compared to the darker regions. Taking Fig.~\ref{fig9b} as an example, the Doppler effect causes the intensity of the direct image on the left side to be greater than that on the right side, resulting in a pronounced distribution of polarization intensity on the left. It is noteworthy that, for black holes, radiation cannot escape beyond the event horizon, so polarization effects theoretically cannot be observed in the inner region. However, strong polarization effects are present in the central region of boson stars, providing an important criterion for distinguishing between the two types of compact objects.

\section{Conclusions and discussions}\label{sec5}
This work investigates the optical appearance of Bardeen-boson stars based on the Einstein-Klein-Gordon theory coupled to NLED, employing the ray-tracing method. Specifically, in a spherically symmetric spacetime, we obtain numerical solutions for the boson star using a shooting method, and further derive an analytical expression for its metric through a fitting procedure. Subsequently, using a thin accretion disk located in the equatorial plane as the light source, we construct the optical and polarized images of the boson star through stereographic projection.

We focus on the influence of the initial scalar field $\phi_0$, the magnetic charge $\mathcal{G}$, and the observer inclination angle $\theta_o$ on the image structure. When the constant $\mathcal{S} = 0.8$ and $\mathcal{G} = 0.35$, the values of $\phi_0 = (0.3, 0.35, 0.55, 0.6)$ correspond to BS1, BS2, BS3, and BS4, respectively. When $\mathcal{S} = 0.2$ and $\phi_0 = 0.6$, the magnetic charge $\mathcal{G} = (0.01, 0.07, 0.1, 0.15)$ corresponds to BS5 through BS8. The numerical results show that as $\phi_0$ and $\mathcal{G}$ increase, the total mass of the boson star decreases. Further fitting results indicate that $g_{tt}$ and $g_{rr}$ approach the Schwarzschild black hole as $r \rightarrow \infty$, exhibiting asymptotic flatness.

Assuming that the material in the thin accretion disk moves along nearly circular orbits, we investigate the influence of $(\phi_0, \mathcal{G}, \theta_o)$ on the optical images of boson stars. The results show that $\theta_o$ primarily determines the shape of the direct image. When $\theta_o = 0^\circ$, the direct image exhibits a standard ring structure; as $\theta_o$ increases to $80^\circ$, the direct image gradually distorts into a cap-like shape. For BS1--BS4, a larger $\phi_0$ significantly reduces the size of the direct image, and under conditions of larger $\phi_0$ and $\theta_o$, D-type lensing images become visible. For BS5--BS8, the increase in $\mathcal{G}$ only slightly reduces the size of the direct image. When $\theta_o = 30^\circ$, lensing images appear in all cases, with their distinguishability improving as $\theta_o$ increases. Notably, regardless of the parameter changes, a brightness depression region always exists at the center of the image, similar to the inner shadow of black holes. This suggests that boson stars have the potential to be "black hole mimickers." To further investigate the possible existence of photon rings in the boson star spacetime, we plot the lensing bands for different values of $\theta_o$. No higher-order images were observed, indicating the absence of photon rings. Additionally, the first derivative of the effective potential $\mathcal{V}_{photon}(r)$ has no zeros in all cases, which further confirms the above conclusion. In contrast, photon rings always exist in black hole spacetimes, providing an important criterion for distinguishing boson stars from black holes.

For the polarized images, the polarization intensity shows a positive correlation with the brightness of the optical image, i.e., the polarization intensity in the high-brightness regions is much higher than in the low-brightness regions. Both the polarization intensity and direction exhibit a strong dependence on $(\phi_0, \mathcal{G}, \theta_o)$, indicating that the polarization characteristics of boson stars can effectively reflect their intrinsic spacetime structure. Interestingly, since boson stars lack an event horizon, polarization effects are present even inside the star. In contrast, for black holes, no polarization effects can be observed inside the event horizon.

This study provides a theoretical basis for distinguishing black holes from boson stars through their optical appearance. Compared to relying solely on accretion disk imaging, we emphasize the importance of combining optical images with polarization effects to obtain more robust and reliable criteria. Future research will focus on boson stars under different potential models and more realistic, complex accretion disk models, which may offer additional insights for distinguishing boson stars from black holes.


\cleardoublepage

\vspace{10pt}
\noindent {\bf Acknowledgments}

\noindent
This work is supported by the National Natural Science Foundation of China (Grants Nos.
12375043, 12575069 ). 

\bibliographystyle{JHEP} 
\bibliography{biblio} 

\end{document}